\documentclass[12pt]{article}
\usepackage{epsfig, epsf, graphicx, subfigure, epstopdf}
\pdfoutput=1
\usepackage{pstricks, pst-node, psfrag}
\usepackage{amssymb,amsmath,bm}
\usepackage{verbatim,enumerate}
\usepackage{rotating, lscape}
\usepackage{setspace}
\usepackage{algorithm}
\usepackage{algorithmicx}
\usepackage{algpseudocode}
\usepackage{natbib}
\usepackage{multirow}

\def\boxit#1{\vbox{\hrule\hbox{\vrule\kern6pt
          \vbox{\kern6pt#1\kern6pt}\kern6pt\vrule}\hrule}}

\def\bse{\begin{eqnarray*}}
\def\ese{\end{eqnarray*}}
\def\be{\begin{eqnarray}}
\def\ee{\end{eqnarray}}
\def\bq{\begin{equation}}
\def\eq{\end{equation}}
\def\bse{\begin{eqnarray*}}
\def\ese{\end{eqnarray*}}

\newcommand{\mbf}{\mathbf}

\newcommand{\iid}{\stackrel{\mathrm{iid}}{\sim}}

\begin{document}

\thispagestyle{empty} \baselineskip=28pt \vskip 5mm
\begin{center} {\Huge{\bf A Stochastic Generator of Global Monthly Wind Energy with Tukey $g$-and-$h$ Autoregressive Processes}}
\end{center}

\baselineskip=12pt \vskip 10mm

\begin{center}\large
Jaehong Jeong\footnote[1]{\baselineskip=10pt CEMSE Division, King Abdullah University of Science and Technology, Thuwal 23955-6900, Saudi Arabia. E-mail: jaehong.jeong@kaust.edu.sa, yuan.yan@kaust.edu.sa, marc.genton@kaust.edu.sa.}, Yuan Yan$^{1}$, Stefano Castruccio\footnote[2]{\baselineskip=10pt Department of Applied and Computational Mathematics and Statistics, 153 Hurley Hall, University of Notre Dame, Notre Dame, IN 46556, USA. E-mail: scastruc@nd.edu.\\ This publication is based upon work supported by the King Abdullah University of Science and Technology (KAUST) Office of Sponsored Research (OSR) under Award No: OSR-2015-CRG4-2640.}, and Marc G.~Genton$^{1}$
\end{center}

%Paola Crippa\footnote[3]{\baselineskip=10pt Department of Civil and Environmental Engineering and Earth Sciences, 156 Fitzpatrick Hall, University of Notre Dame, Notre Dame, IN 46556, USA. E-mail: pcrippa@nd.edu.

%\baselineskip=17pt \vskip 10mm \centerline{\today} \vskip 10mm
\baselineskip=17pt \vskip 10mm \centerline{November 10, 2017} \vskip 10mm

%%%%%%%%%%%%%%%%%%%%%%%%%%%%%%%%%%%%%%%%%%%%%%%%%%%%%%%%%%%%%%%%%%%%%%%%
\begin{center}
{\large{\bf Abstract}}
\end{center}

\baselineskip=17pt

Quantifying the uncertainty of wind energy potential from climate models is a very time-consuming task and requires a considerable amount of computational resources. A statistical model trained on a small set of runs can act as a stochastic approximation of the original climate model, and be used to assess the uncertainty considerably faster than by resorting to the original climate model for additional runs. While Gaussian models have been widely employed as means to approximate climate simulations, the Gaussianity assumption is not suitable for winds at policy-relevant time scales, i.e., sub-annual. We propose a trans-Gaussian model for monthly wind speed that relies on an autoregressive structure with Tukey $g$-and-$h$ transformation, a flexible new class that can separately model skewness and tail behavior. This temporal structure is integrated into a multi-step spectral framework that is able to account for global nonstationarities across land/ocean boundaries, as well as across mountain ranges. Inference can be achieved by balancing memory storage and distributed computation for a data set of 220 million points. Once fitted with as few as five runs, the statistical model can generate surrogates fast and efficiently on a simple laptop, and provide uncertainty assessments very close to those obtained from all the available climate simulations (forty) on a monthly scale.

\par\vfill\noindent
\par\medskip\noindent
{\bf Short title}: Stochastic Monthly Wind Generators

\clearpage\pagebreak\newpage \pagenumbering{arabic}
\baselineskip=27pt

%%%%%%%%%%%%%%%%%%%%%%%%%%%%%%%%%%%%%%%%%%%%%%%%%%%%%%%%%%%%%%%%%%%%%%%%
\section{Introduction}\label{sec:intro}
%%%%%%%%%%%%%%%%%%%%%%%%%%%%%%%%%%%%%%%%%%%%%%%%%%%%%%%%%%%%%%%%%%%%%%%%

Wind energy plays an important role in many countries' energy portfolio as a significant renewable source with no major negative environmental impacts \citep{wiser2011wind,Obamaaam6284}. Earth System Models (ESMs) provide physically consistent projections of wind energy potential, as well as spatially resolved maps in regions with poor observational coverage. However, these models are (more or less accurate) approximations of the actual state of the Earth's system and the energy assessment is therefore sensitive to changes in the model input. To address this, geoscientists generate a collection (\textit{ensemble}) of ESMs to assess the sensitivity of the output (including wind) with respect to physical parameters and trajectories of greenhouse gases concentration (forcing scenarios). Recently, the role of the uncertainty due to ESMs' initial conditions (\textit{internal variability}) has been identified as a prominent factor for multi-decadal projections, hence the importance of quantifying its uncertainty. 

The Large ENSemble (LENS) is a collection of 40 runs at the National Center for Atmospheric Research (NCAR) specifically designed to isolate the role of internal variability in the future climate \citep{kay2015community}. The LENS required millions of CPU hours on a specialized supercomputer, and very few institutions have the resources and time for such an investigation. Is such an enormous task always necessary for assessing internal variability? While it is absolutely necessary for quantities at the tail of the climate (e.g., temperature extremes), it is not always necessary for simpler indicators such as climate mean and variance. As part of a series of investigations promoted by KAUST on the topic of assessing wind energy in Saudi Arabia, \cite{jeong2017reducing} introduced the notion of a stochastic generator (SG), a statistical model that is trained on a small subset of LENS runs. The SG acts as a stochastic approximation of the climate model and hence allows for sampling more surrogate climate runs\footnote{A brief discussion on the difference between a SG and an emulator is contained in the same work.}. In their study, the authors present a SG for the global annual wind and show that only five runs are sufficient to generate synthetic runs visually indistinguishable from the original simulations, and with a similar spatio-temporal local dependence. However, while the SG introduced by the authors is able to approximate annual global data for the Arabian peninsula effectively, an annual scale is not useful for wind energy assessment, and a SG at a finer temporal resolution in the same region is required to provide policy-relevant results. 

A SG for monthly global wind output requires considerable modeling and computational efforts. From a modeling perspective, data indexed on the sphere and in time require a dependence structure able to incorporate complex nonstationarities across the entire Earth's system, see \cite{jeong2017global} for a recent review of multiple approaches. For regularly spaced data, as is the case with atmospheric variables in an ESM output, multi-step spectrum models are particularly useful as they can provide flexible nonstationary structures for Gaussian processes in the spectral domain while maintaining positive definiteness of the covariance functions \citep{castruccio2013global,castruccio2014beyond,cas16,castruccio2016compressing,cas17b}. Recently, \cite{castruccio2017evolutionary} and \cite{jeong2017reducing} introduced a generalization that allows graphical descriptors such as land/ocean indicators and mountain ranges to be incorporated in a spatially varying spectrum. 

Besides the modeling complexity, the computational challenges are remarkable, as inference needs to be performed on an extremely large data set. Over the last two decades, the increase in size of spatio-temporal data sets in climate has prompted the development of many new classes of scalable models. Among the many solutions proposed, fixed rank methods \citep{cre08}, predictive processes \citep{banerjee2008gaussian}, covariance tapering \citep{fue06}, and Gaussian Markov random fields \citep{rue05} have played a key role in our ability to couple feasibility of the inference while retaining essential information to be communicated to stakeholders; see \cite{sun2012geostatistics} for a review. However, even by modern spatio-temporal data set standards, 220 million points is a considerable size, and to perform inference, a methodology that leverages on both the parallel computing and gridded geometry of the data is absolutely necessary. \cite{castruccio2013global}, \cite{castruccio2014beyond}, \cite{cas16}, \cite{castruccio2016compressing}, \cite{castruccio2017evolutionary}, \cite{cas17b}, and \cite{jeong2017reducing} have provided a framework for a fast and parallel methodology for extremely large climate data sets. However, it has so far been limited to Gaussian processes. Whether an extension to non-Gaussian models with such a large data set is possible (and how) was an open question. 

In this paper, we propose a SG for monthly winds that is multi-step, spectral and can capture a non-Gaussian behavior. We adopt a simple yet flexible approach to construct non-Gaussian processes in time: the Tukey $g$-and-$h$ autoregressive process \citep{yan2017tukey}, defined as $Y(t) = \xi +\omega \tau_{g,h}\{ Z(t)\}$, where $\xi$ is a location parameter, $\omega$ is a scale parameter, $Z(t)$ is a Gaussian autoregressive process, and $\tau_{g,h}(z)$ is the Tukey $g$-and-$h$ transformation \citep{Tukey77}:  
\be\label{tukey_eq}
\tau_{g,h}(z)=\begin{cases}
g^{-1}\{\exp(gz)-1\}\exp(hz^{2}/2) &\mbox{if $g\neq 0$}, \\
z\exp(hz^{2}/2) & \mbox{if $g=0$},
\end{cases} 
\ee
where $g$ controls the skewness and $h$ governs the tail behavior. A significant advantage of Tukey $g$-and-$h$ autoregressive processes is that they provide very flexible marginal distributions, allowing skewness and heavy tails to be adjusted. This class of non-Gaussian processes is integrated within the multi-step spectral scheme to still allow inference for a very large data set. The Tukey $g$-and-$h$ transformation has also been successfully implemented for max-stable processes \citep{xu2016tukey} and random fields \citep{xu2017tukey}; see \cite{xu2015efficient} and references therein for the independent case. 

The remainder of the paper is organized as follows. Section~\ref{sec:data} describes the wind data set. Section~\ref{sec:covariance} details the statistical framework with the Tukey $g$-and-$h$ autoregressive models and the inferential approach. Section~\ref{sec:comparison} provides a model comparison and Section~\ref{sec:simul} illustrates how to generate SG runs. The article ends with concluding remarks in Section~\ref{sec:concl}.

%%%%%%%%%%%%%%%%%%%%%%%%%%%%%%%%%%%%%%%%%%%%%%%%%%%%%%%%%%%%%%%%%%%%%%%%
\section{The Community Earth System Model (CESM) Large ENSemble project (LENS)}\label{sec:data}
%%%%%%%%%%%%%%%%%%%%%%%%%%%%%%%%%%%%%%%%%%%%%%%%%%%%%%%%%%%%%%%%%%%%%%%%

\begin{figure}[t!]\centering
\includegraphics[height=2in,width=3.25in]{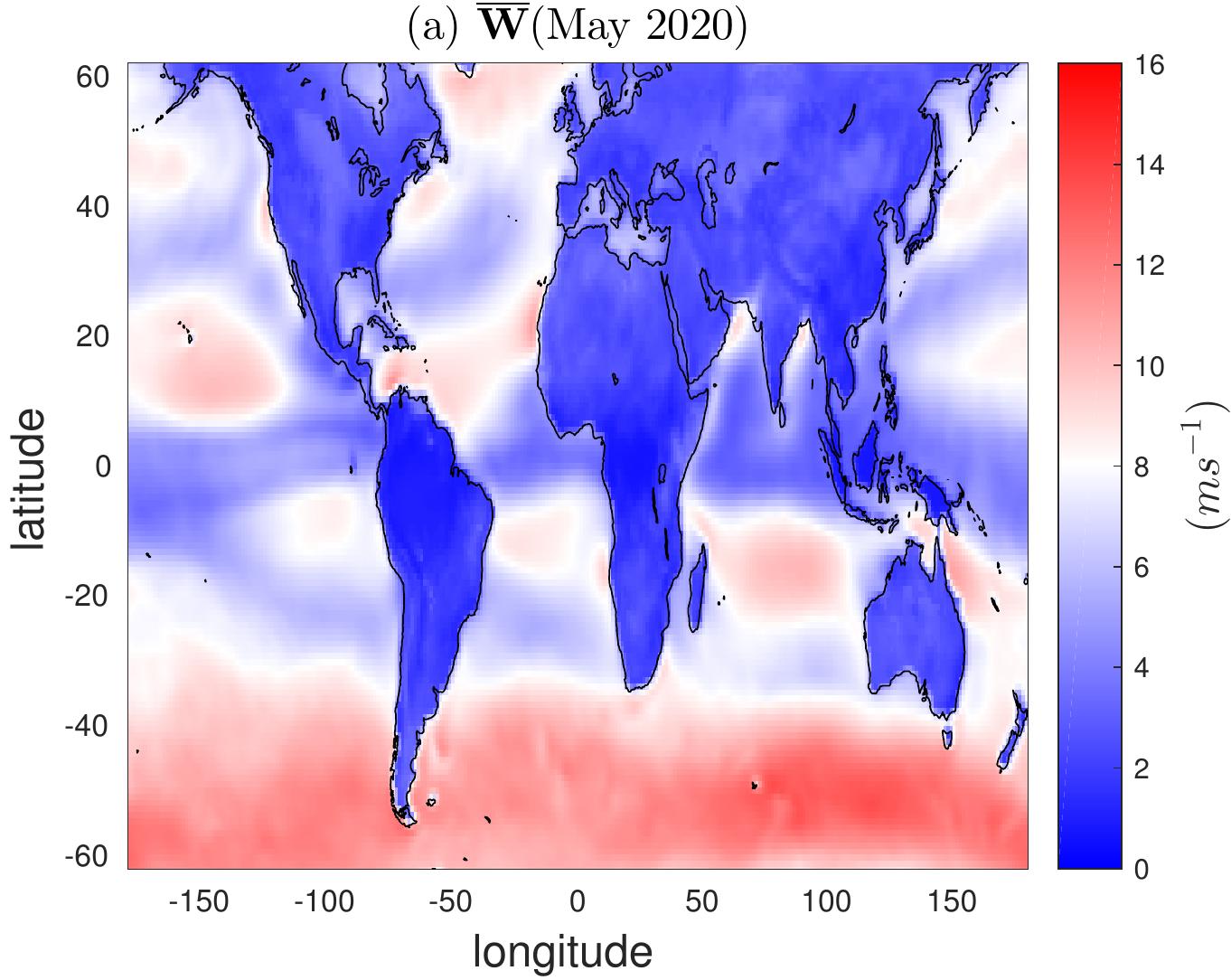}\hspace{0.2cm}
\includegraphics[height=2in,width=3.25in]{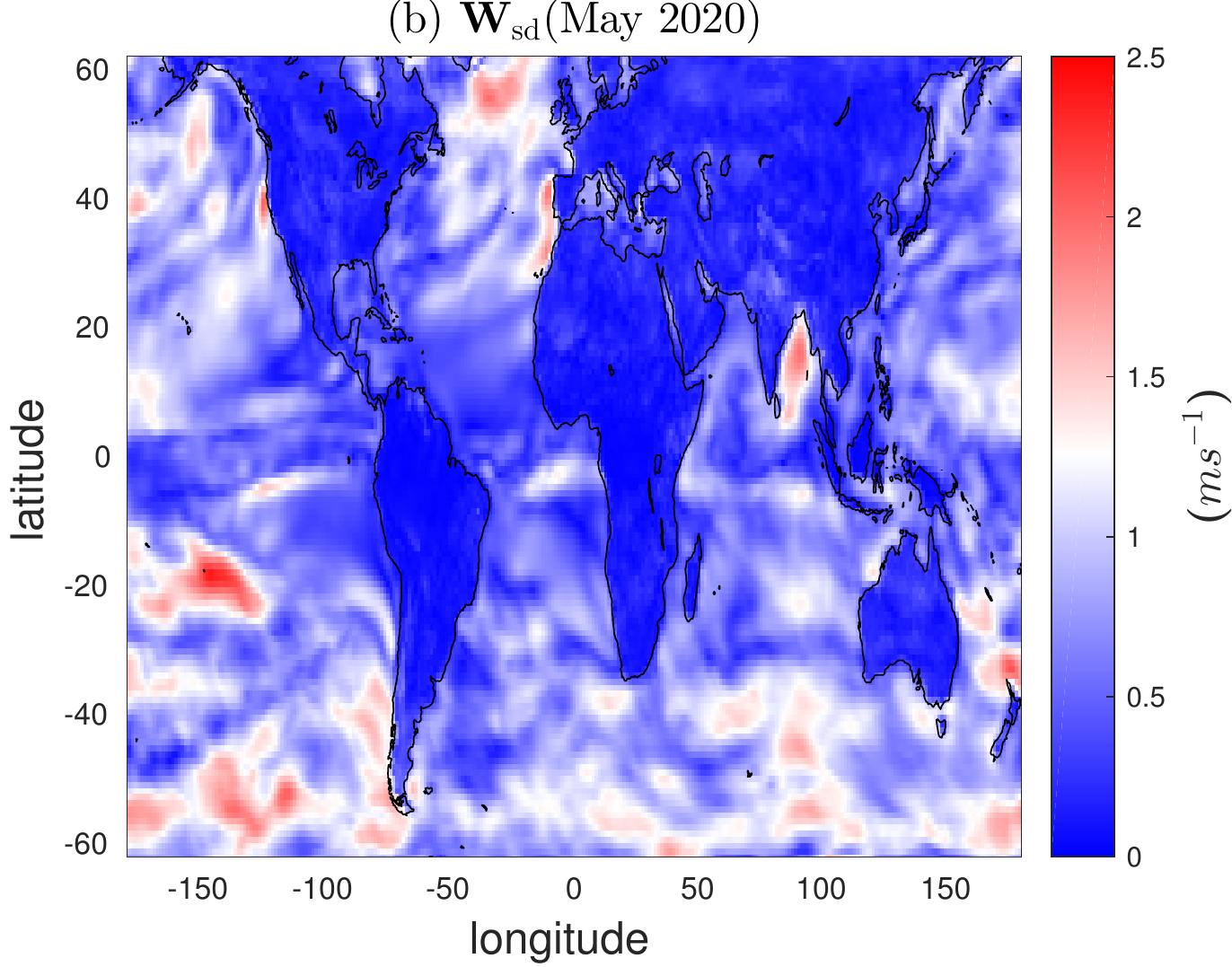}\vspace{0.2cm}
\includegraphics[height=2in,width=3.25in]{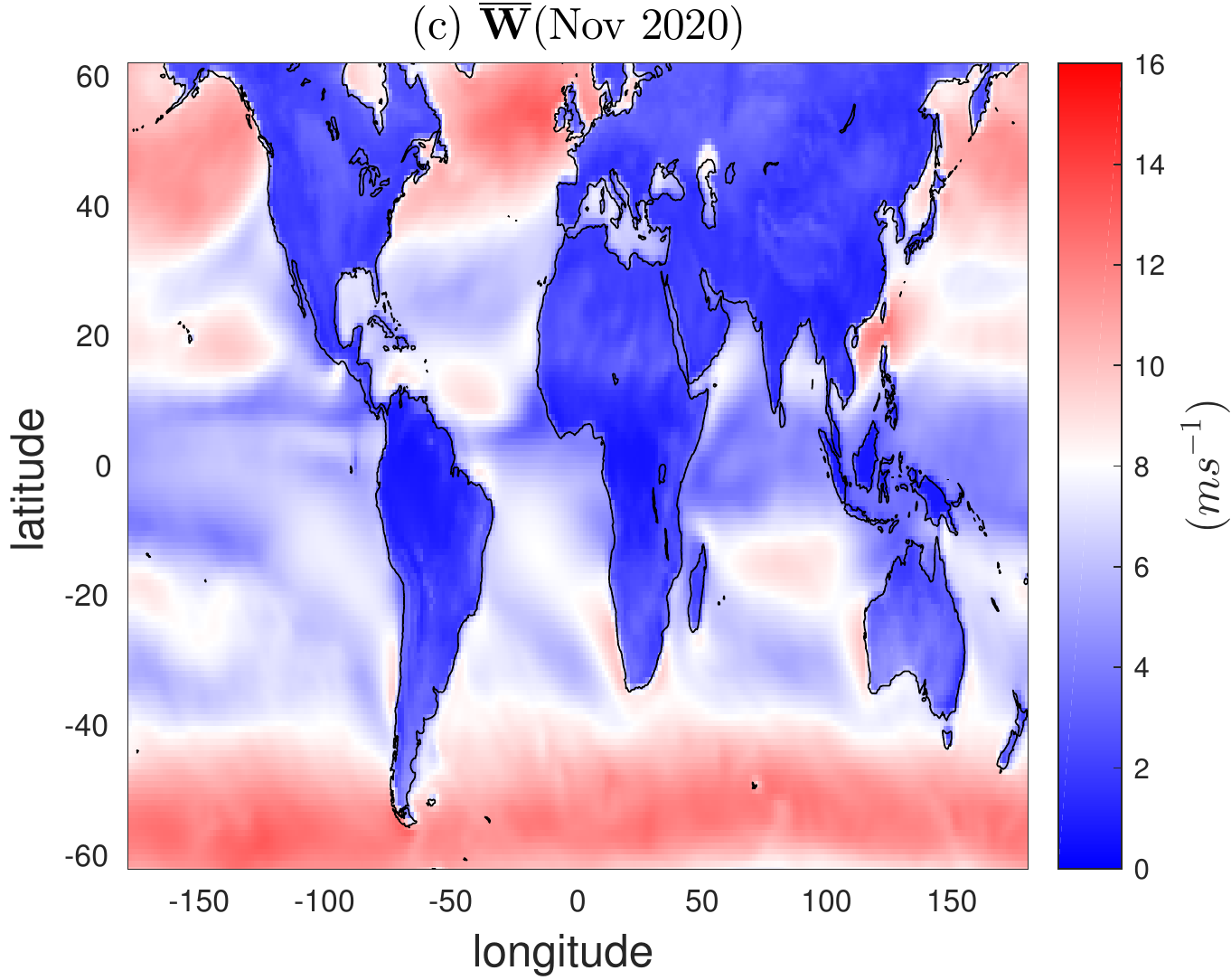}\hspace{0.2cm}
\includegraphics[height=2in,width=3.25in]{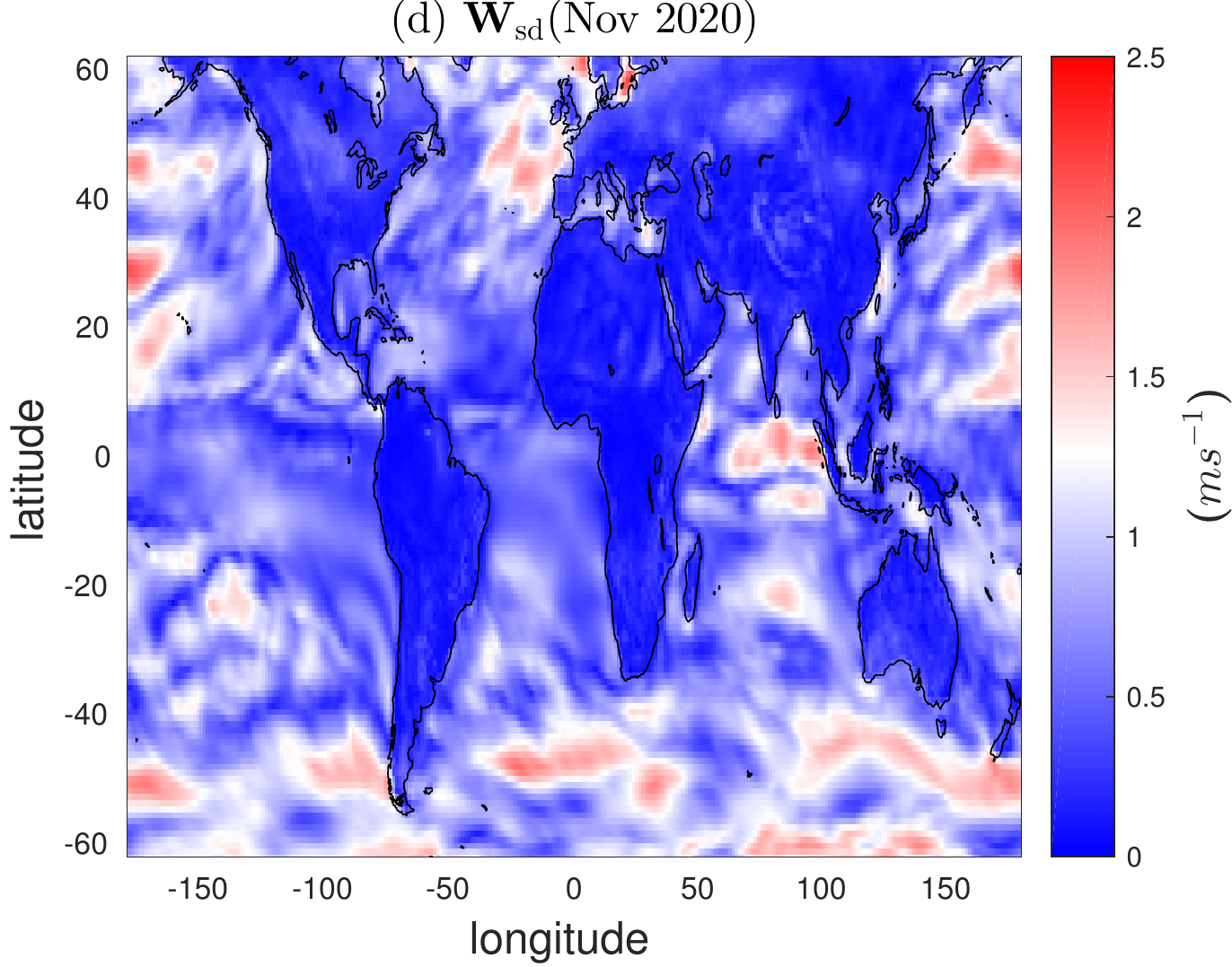}\vspace{0.2cm} 
\includegraphics[height=2in,width=3.15in]{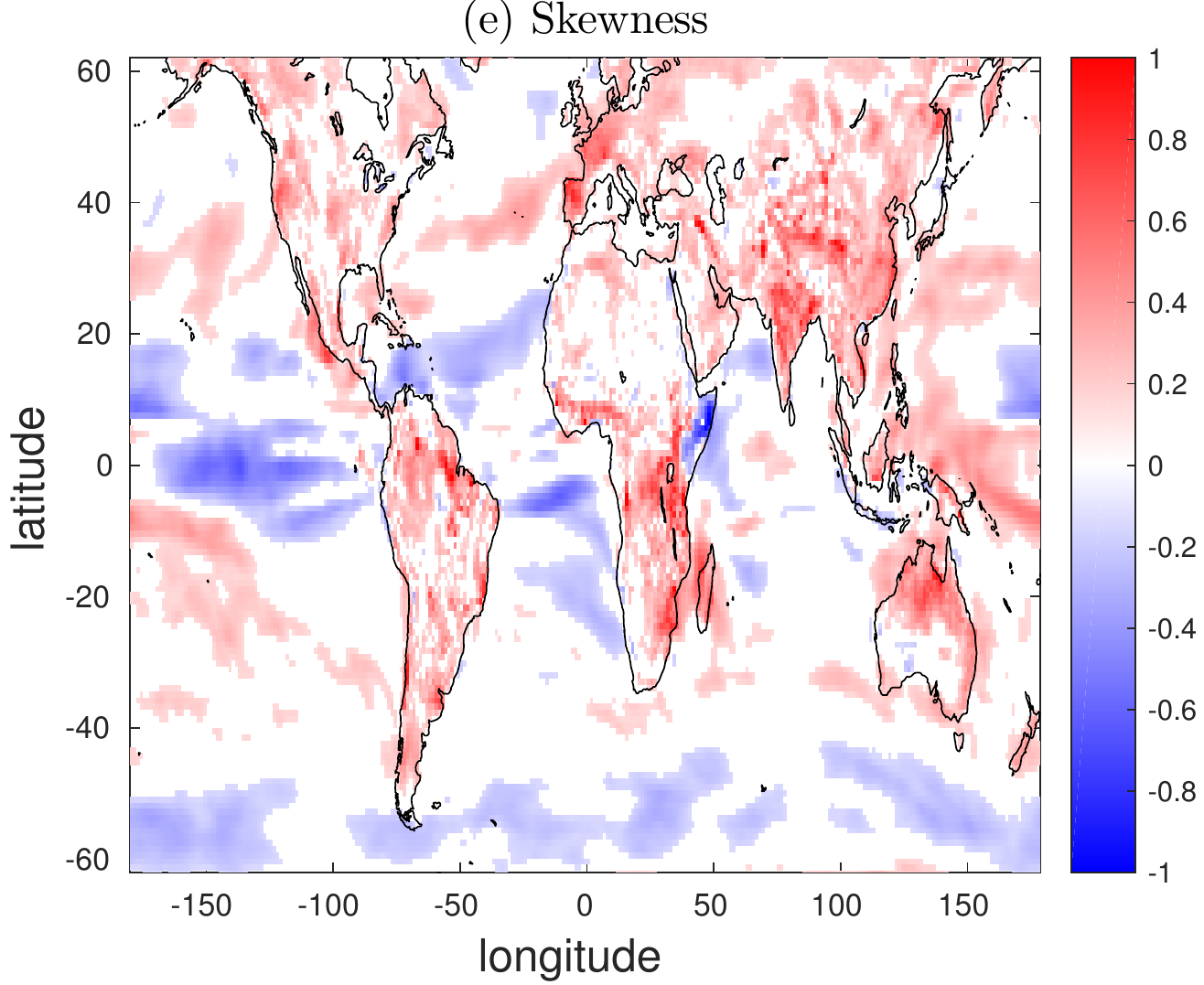}\hspace{0.5cm}
\includegraphics[height=2in,width=3in]{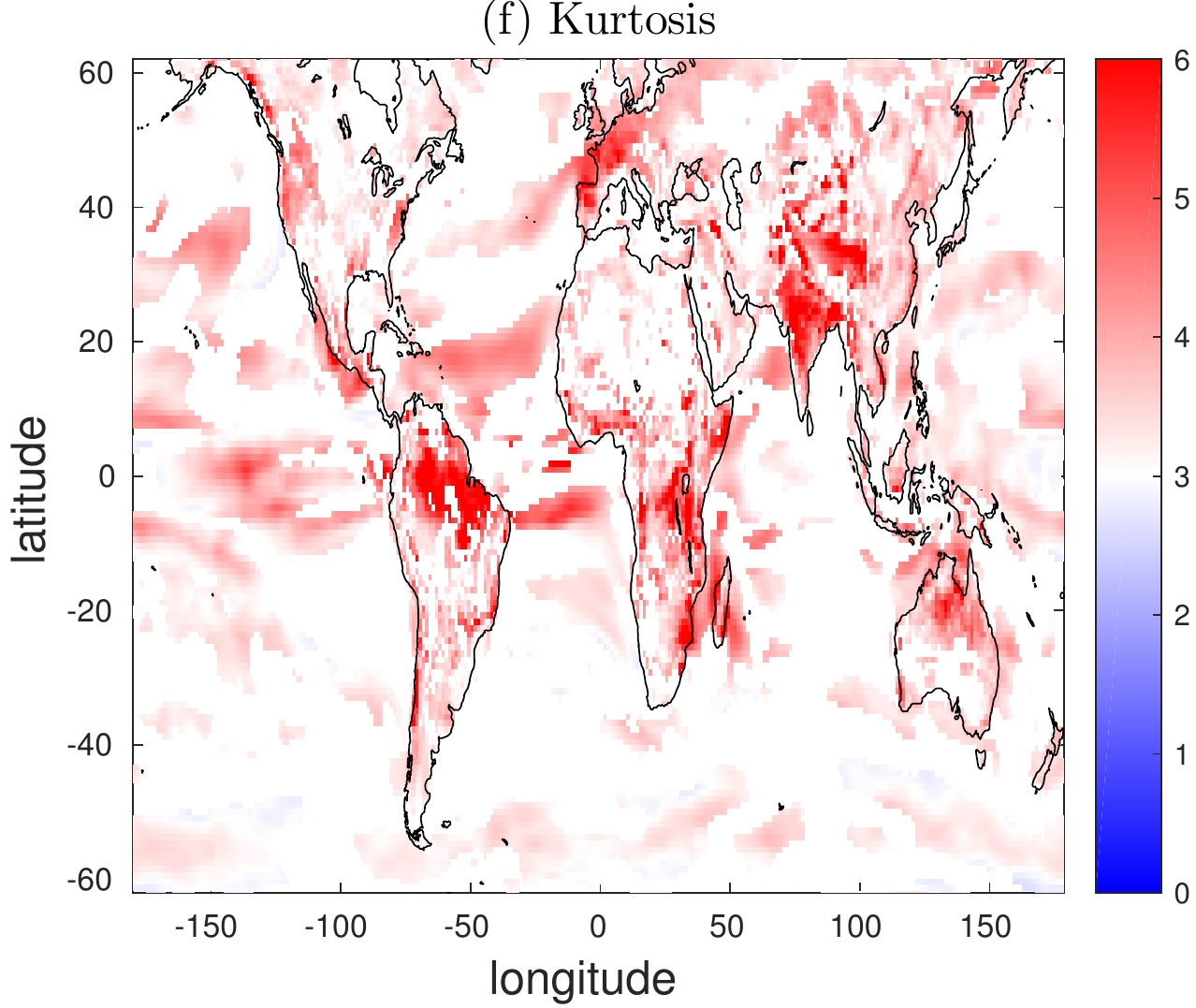}\hspace{0.6cm} 
\caption{The (a) ensemble mean $\overline{\bf W}({\rm May}\ 2020)=\sum_{r=1}^{R}{\bf W}_{r}({\rm May}\ 2020)/R$ where $R=5$ is the number of ensemble members and (b) ensemble standard deviation ${\bf W}_{\rm sd}({\rm May}\ 2020)=\sqrt{\sum_{r=1}^{R}\{{\bf W}_{r}({\rm May}\ 2020)-\overline{\bf W}({\rm May}\ 2020)\}^{2} /R}$, of the monthly wind speed (in $m s^{-1}$). (c) and (d) are the same as (a) and (b), but those in November 2020. The empirical skewness and kurtosis of wind speed from one ensemble member after removing the trend are reported in (e) and (f), respectively, only for the locations where p$\textendash$values of a significance test are less than 0.05.}\label{fig:statistics}
\end{figure}

We work on global wind data from LENS, which is an ensemble of CESM runs with version 5.2 of the Community Atmosphere Model from NCAR \citep{kay2015community}. The ensemble comprises runs at $0.9375^{\circ}\times 1.25^{\circ}$ (latitude $\times$ longitude) resolution, with each run under the Representative Concentration Pathway (RCP) 8.5 \citep{vanvuuren11}. Although the full ensemble consists of 40 runs, in our training set we consider only $R=5$ randomly chosen runs for the SG to demonstrate that only a small number of runs is necessary (a full sensitivity analysis for $R$ is performed in \citealp{jeong2017reducing}). We consider monthly near-surface wind speed at 10~m above the ground level (U10 variable) from 2006 to 2100. Since our focus is on future wind trends, we analyze the projections for a total of 95 years. We consider all 288 longitudes, and we discard latitudes near the poles to avoid numerical instabilities, due to the very close physical distance of neighboring points and the very different statistical behavior of wind speed in the Arctic and Antarctic regions \citep{mcinnes2011global}, and consistently with previous works. Therefore, we use 134 bands between $62^{\circ}$S and $62^{\circ}$N, and the full data set comprises approximately 220 million points ($5 \times 1140 \times 134\times 288$). An example is given in Figure~\ref{fig:statistics}(a-d) where we show the ensemble mean and standard deviation of the monthly wind speed from the five selected runs, in May and November 2020. We observe that both means and standard deviations show temporal patterns. In particular, between the Northern Tropic and latitude $60^{\circ}$N, the mean of wind speed over the ocean in November is stronger than that in May.

For each site, we test the significance of skewness and kurtosis of wind speed over time \citep{bai2005tests} after removing the climate. In many spatial locations, the p$\textendash$values are smaller than 0.05 as shown in Figure~\ref{fig:statistics}(e) and (f), thus indicating how the first two moments are not sufficient to characterize the temporal behavior of monthly wind in time. Most land points have significant skewness and, consistently with \cite{bauer1996characteristic}, we observe that monthly wind speeds over the ocean are negatively skewed in the tropics and positively skewed outside of that region. The Tropical Indian Ocean and the Western Pacific Ocean, both areas of small wind speeds, represent an exception of positively skewed distribution.

%%%%%%%%%%%%%%%%%%%%%%%%%%%%%%%%%%%%%%%%%%%%%%%%%%%%%%%%%%%%%%%%%%%%%%%%
\section{The Space-Time Model}\label{sec:covariance}
%%%%%%%%%%%%%%%%%%%%%%%%%%%%%%%%%%%%%%%%%%%%%%%%%%%%%%%%%%%%%%%%%%%%%%%%

%%%%%%%%%%%%%%%%%%%%%%%%%%%%%%%%%%%%%%%%%%%%%%%%%%%%%%%%%%%%%%%%%%%%%%%%
\subsection{The Statistical Framework}\label{sec:newM}

It is known that, after the climate model forgets its initial state, each ensemble member evolves in `deterministically chaotic' patterns \citep{lorenz1963deterministic}. Climate variables in the atmospheric module have a tendency to forget their initial conditions after a short period, and to evolve randomly while still being attracted by the mean climate. Since ensemble members from the LENS differ only in their initial conditions \citep{kay2015community}, we treat each one as a statistical realization from a common distribution in this work. We define ${W}_{r}(L_{m},\ell_{n},t_{k})$ as the spatio-temporal monthly wind speed for realization $r$ at the latitude $L_{m}$, longitude $\ell_{n}$, and time $t_{k}$, where $r=1,\dots,R$, $m=1,\dots,M$, $n=1,\dots,N$, and $k=1,\dots,K$, and define ${\bf W}_{r}=\{{W}_{r}(L_{1},\ell_{1},t_{1}),\dots,{W}_{r}(L_{M},\ell_{N},t_{K})  \}^{\top}$.

To remove the trend in our model, we consider ${\bf D}_{r}={\bf W}_{r}-\overline{\bf W}$ with $\overline{\bf W}=\frac{1}{R}\sum_{r=1}^R {\bf W}_{r}$. The Gaussian assumption for ${\bf D}_{r}$ is not in general valid at monthly resolution (see Figure \ref{fig:statistics}(e-f), Figure~S1 for a significance test on the skewness and kurtosis, and Figure~S2 for Lilliefors and Jarque-Bera normality tests); we therefore apply the Tukey $g$-and-$h$ transformation \eqref{tukey_eq}, so that our model can be written as:
\be\label{eq_modelT}
{\bf D}_{r}=\underline{{\bm \xi}}+\underline{\bm \omega} \cdot {\bm \tau_{\underline{\bm g},\underline{\bm h}}}({\bm \epsilon}_{r}),\quad {\bm \epsilon}_{r}\stackrel{\rm iid}{\sim}\mathcal{N}({\bm 0},{\bm \Sigma}({\bm \theta}_{\rm space-time})),
\ee
where $\underline{{\bm \xi}}={\bm \xi} \otimes \mathbf{1}_K$, with ${\bm \xi}=\{\xi(L_1,\ell_1),\ldots, \xi(L_M,\ell_N)\}^{\top}$ being the vector of the location parameters, $\otimes$ the Kroneker product, $\mathbf{1}_K$ the vector of ones of length $K$, $\underline{{\bm \omega}}={\bm \omega} \otimes \mathbf{1}_K$, with ${\bm \omega}=\{\omega(L_1,\ell_1),\ldots, \omega(L_M,\ell_N)\}^{\top}$ the vector of scale parameters, $\underline{{\bm g}}={\bm g} \otimes \mathbf{1}_K$, and $\underline{{\bm h}}={\bm h} \otimes \mathbf{1}_K$, with ${\bm g}=\{g(L_1,\ell_1),\ldots, g(L_M,\ell_N)\}^{\top}$ and ${\bm h}=\{h(L_1,\ell_1),\ldots, h(L_M,\ell_N)\}^{\top}$ the vectors of the $MN$ parameters for the Tukey $g$-and-$h$ transformation at each site. Here ${\bm \tau_{\underline{\bm g},\underline{\bm h}}}()$ represents the element-wise transformation according to \eqref{tukey_eq}. 

We denote by $\bm{\theta}_{\rm Tukey}=({\bm \xi}^{\top},{\bm \omega}^{\top},{\bm g}^{\top},{\bm h}^{\top})^{\top}$ the parameters of the Tukey $g$-and-$h$ transformation and by ${\bm \theta}_{\rm space-time}=(\bm{\theta}_{\rm time}^{\top},\bm{\theta}_{\rm lon}^{\top},\bm{\theta}_{\rm lat}^{\top})^{\top}$ the vector of covariance parameters, which can be divided into temporal, longitudinal, and latitudinal dependence. The total set of parameters is ${\bm \theta}=(\bm{\theta}_{\rm Tukey}^{\top},{\bm \theta}_{\rm space-time}^{\top})^{\top}$.

Here ${\bm \theta}$ is very high dimensional, hence we consider a multi-step inference scheme as first introduced by \cite{castruccio2013global}, where the parameters obtained from previous steps are assumed fixed and known:

\begin{itemize}
\itemsep0em
\item[] Step 1. We estimate $\bm{\theta}_{\rm Tukey}$ and $\bm{\theta}_{\rm time}$ by assuming that there is no cross-temporal dependence in latitude and longitude;
\item[] Step 2. We consider $\bm{\theta}_{\rm Tukey}$ and $\bm{\theta}_{\rm time}$  fixed at their estimated values and estimate $\bm{\theta}_{\rm lon}$ by assuming that the latitudinal bands are independent;
\item[] Step 3. We consider $\bm{\theta}_{\rm Tukey},\bm{\theta}_{\rm time}$ and $\bm{\theta}_{\rm lon}$ fixed at their estimated values and estimate $\bm{\theta}_{\rm lat}$. 
\end{itemize}

This conditional step-wise approach implies some degree of error and uncertainty propagation across stages. \cite{castruccio2017evolutionary} provided some guidelines on how to control for the propagation by using intermediate steps within Step 3. Following the same scheme, we detail the model for each of the three steps and provide a description for the inference (see Sections~\ref{first_step}, \ref{second_step}, and \ref{third_step}).

%more here on uncertainty propagation? 

%Step 1 estimates $\bm{\theta}_{\rm tukey}$ and $\bm{\theta}_{\rm time}$ for each realization and consider the average across the fit {\color{red}STEFANO: please check if it is correct. They were used separately to make innovations}. Steps 2 and 3, however, are performed wit a Restricted Maximum likelihood approach (REML), see the following sections.  

%%%%%%%%%%%%%%%%%%%%%%%%%%%%%%%%%%%%%%%%%%%%%%%%%%%%%%%%%%%%%%%%%%%%%%%%

\subsection{Step 1: Temporal Dependence and Inference for the Tukey $g$-and-$h$ model}\label{first_step}

We assume that ${\bm \epsilon}_{r}=\{{\bm \epsilon}_{r}(t_1)^{\top},\ldots, {\bm \epsilon}_{r}(t_K)^{\top}\}^{\top}$ in \eqref{eq_modelT} evolves according to a Vector AutoRegressive model of order $p$ (VAR($p$)) with different parameters for each spatial location:
\be\label{eq:tar}
{\bm \epsilon}_{r}(t_{k})={\bm \Phi}_{1}{\bm \epsilon}_{r}(t_{k-1})+\cdots+{\bm \Phi}_{p}{\bm \epsilon}_{r}(t_{k-p})+{\bf S}{\bf H}_{r}(t_{k}), \quad {\bf H}_{r}(t_{k})\iid \mathcal{N}(\mbf{0},\mbf{C}(\bm{\theta}_{\rm lon},\bm{\theta}_{\rm lat})),
\ee
where ${\bm \Phi}_{1}=\text{diag}\{\phi_{L_{m},\ell_{n}}^{1}\}, \ldots, {\bm \Phi}_{p}=\text{diag}\{\phi_{L_{m},\ell_{n}}^{p}\}$ are $MN\times MN$ diagonal matrices with autoregressive coefficients, and  ${\bf S}=\text{diag}\{S_{L_{m},\ell_{n}}\}$ is the diagonal matrix of the standard deviations. Such a model assumes that there is no cross-temporal dependence across locations, and Figure~S3(a) in the supplementary material provides diagnostics on this assumption. Hence, at each spatial location, we have a Tukey $g$-and-$h$ autoregressive process of order $p$ \citep{yan2017tukey}. The vector of temporal parameters is therefore ${\bm \theta}_{\rm time}=(\phi^{1}_{L_{m},\ell_{n}},\ldots,\phi^{p}_{L_{m},\ell_{n}}, S_{L_{m},\ell_{n}})^{\top}, n=1,\dots,N$ and $m=1,\dots,M$.

The vectors $\bm{\theta}_{\rm Tukey}$ and ${\bm \theta}_{\rm time}$ in \eqref{eq_modelT} are estimated simultaneously via a maximum approximated likelihood estimation (MALE, \cite{xu2015efficient}), and since the model assumes no cross-temporal dependence, each site can be analyzed independently and it is straightforward to parallelize the inference. For each spatial location, we select a model and compute the optimal order for \eqref{eq:tar}. The results are shown in Figure~S3(b). For a substantial share of points ($56.1\%$), $p>0$ was selected, hence underscoring the need for a model with temporal dependence, even after differencing the original data from the average across realizations. A map of ${\hat{\phi}_{L_{m},\ell_{n}}^{1}}, {\hat{\phi}_{L_{m},\ell_{n}}^{2}}$ and ${\hat{\phi}_{L_{m},\ell_{n}}^{3}}$ is shown in the supplementary material (Figure~S5), along with the p$\textendash$values (Figure~S6).

\begin{figure}[t!]\centering
\includegraphics[height=2in,width=3.2in]{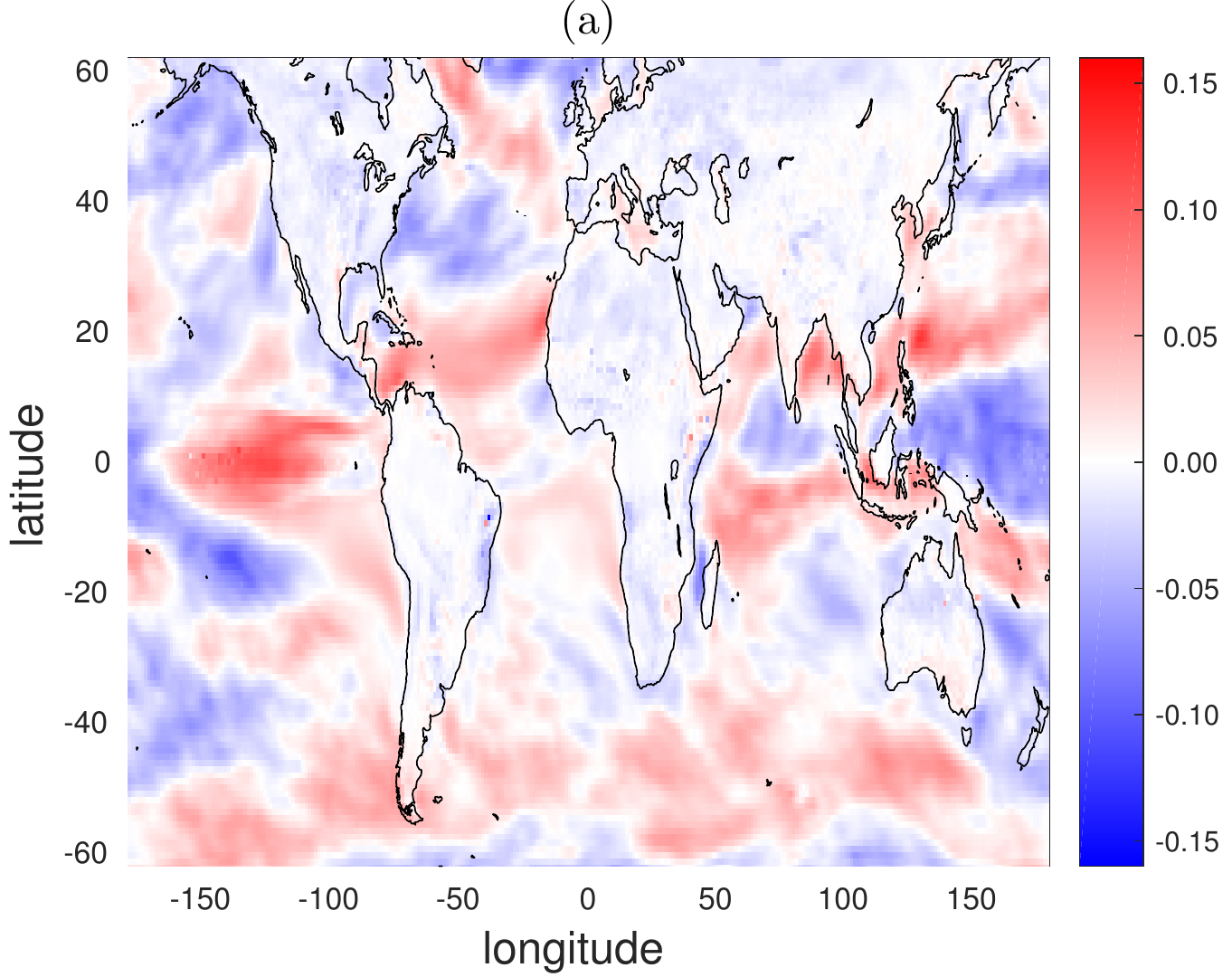}\hspace{0.2cm}
\includegraphics[height=2in,width=3.2in]{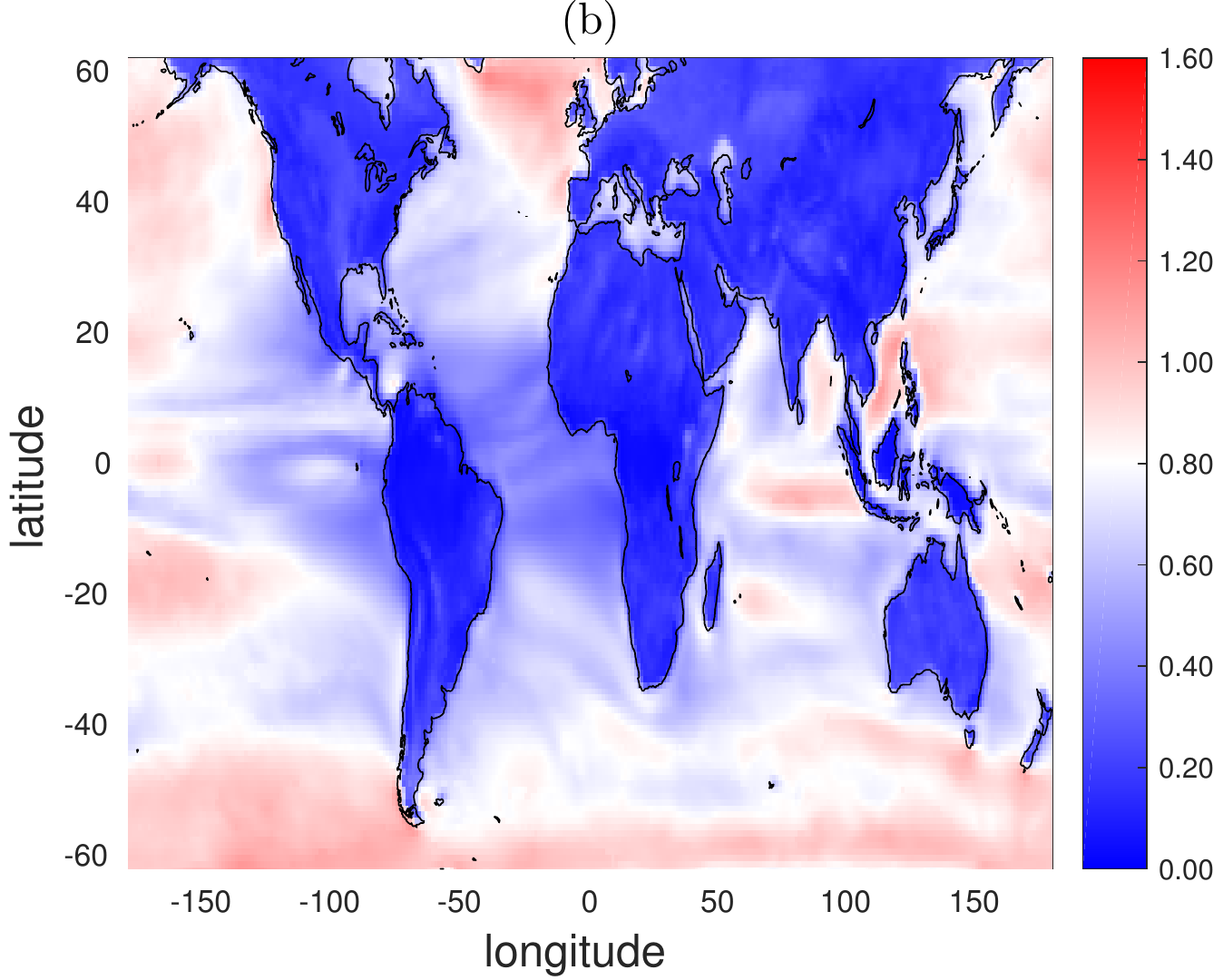}\vspace{0.2cm}
\includegraphics[height=2in,width=3.2in]{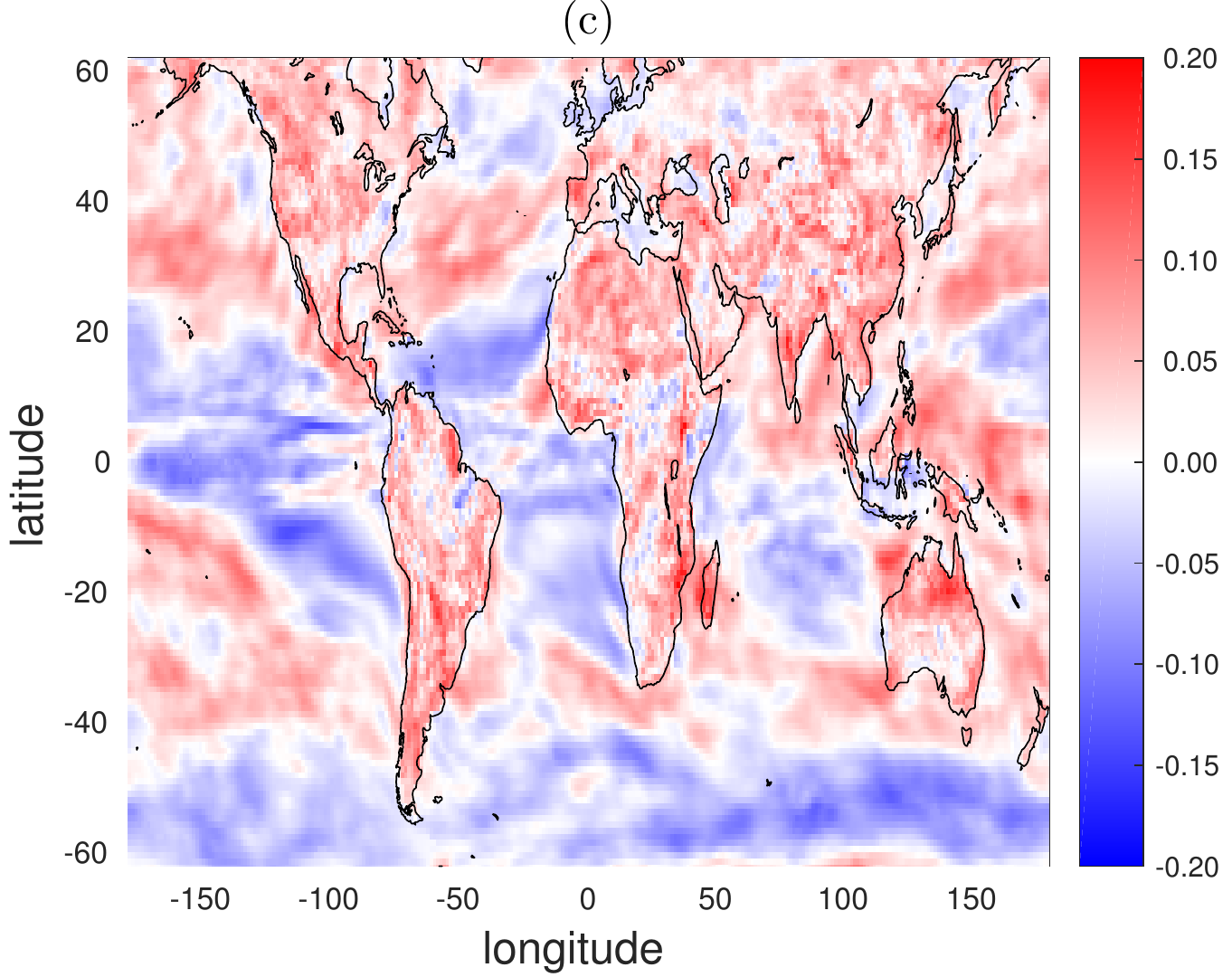}\hspace{0.2cm}
\includegraphics[height=2in,width=3.2in]{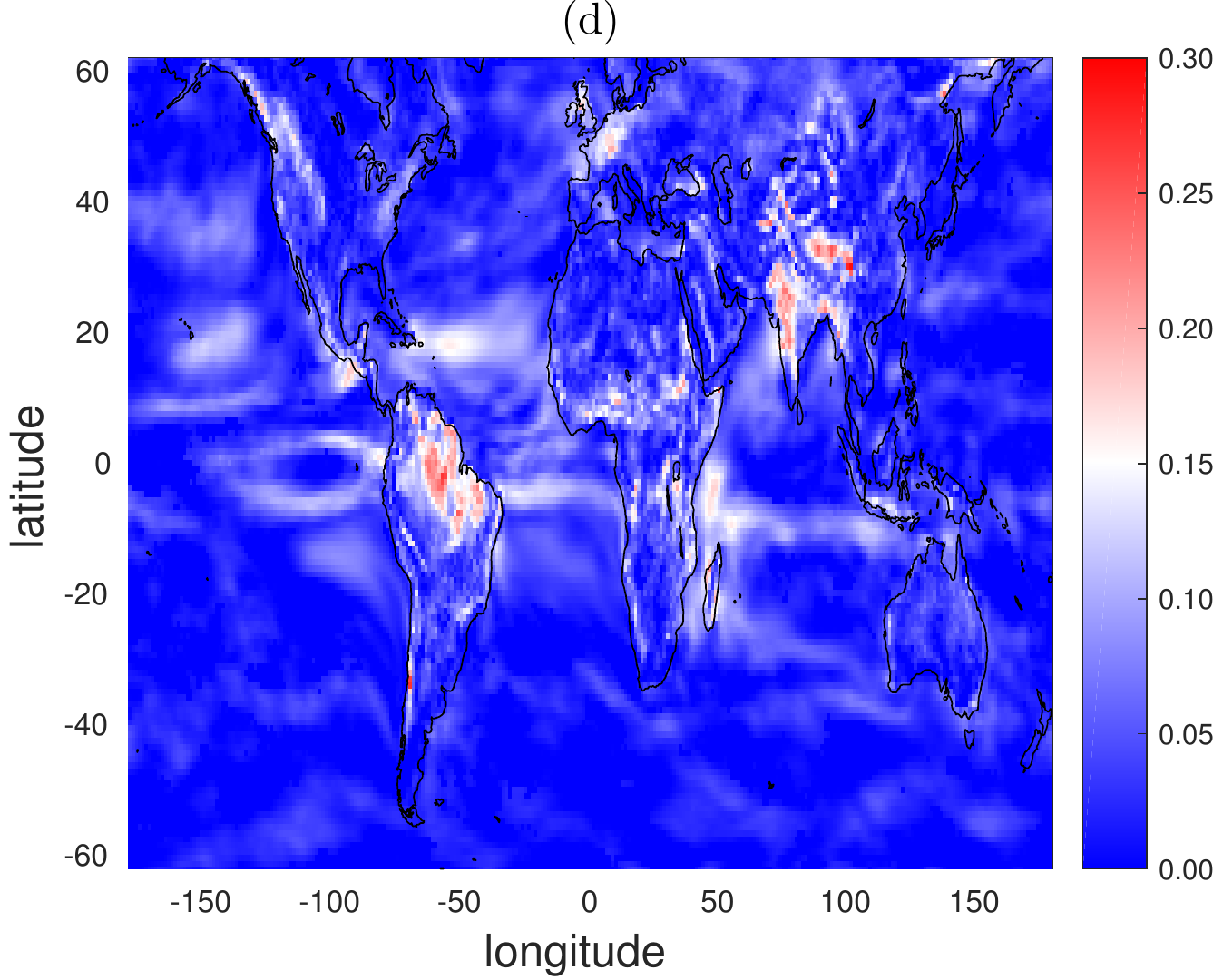}
\caption{Plot of the estimated parameters ${\hat {\boldsymbol \theta}}_{\rm Tukey}$ for the Tukey $g$-and-$h$ transformation, (a) location, (b) scale, (c) $g$, and (d) $h$.}\label{fig:Tukey_est}
\end{figure}

Estimated values $\hat{\bm{\theta}}_{\rm Tukey}$ are represented in Figure~\ref{fig:Tukey_est}. Here $\hat{g}(L_m,\ell_n)$ and $\hat{h}(L_m,\ell_n)$ were estimated with significant non-zero values over many locations (see Figure~S4 for the p$\textendash$value), and it is apparent how the Gaussian autoregressive model is not suitable for modeling monthly wind speed. 

Once all parameters are estimated, the residuals can be calculated as,
\be
{\hat{H}}_{r}(L_{m},\ell_{n},t_{k})=\frac{1}{\hat{S}(L_m,\ell_n)}\left\{\hat{\epsilon}_{r}(L_{m},\ell_{n},t_{k})-\hat{\phi}^{1}_{L_{m},\ell_{n}} \hat{\epsilon}_{r}(L_{m},\ell_{n},t_{k-1})-\cdots - \hat{\phi}^{p}_{L_{m},\ell_{n}} \hat{\epsilon}_{r}(L_{m},\ell_{n},t_{k-p})\right\}, \label{eq:innovation}
\ee
where ${\hat{\epsilon}}_{r}(L_{m},\ell_{n},t_{k})=\hat{\tau}_{\hat{g}(L_{m},\ell_{n}),\hat{h}(L_{m},\ell_{n})}^{-1} [\{{{D}}_{r}(L_{m},\ell_{n},t_{k}) -\hat{\xi}(L_{m},\ell_{n})\}/\hat{\omega}(L_{m},\ell_{n})]$, and $\hat{\tau}_{\hat{g}(L_{m},\ell_{n}),\hat{h}(L_{m},\ell_{n})}^{-1}$ are the inverse Tukey $g$-and-$h$ transformations at latitude $L_m$ and longitude $\ell_n$. 

The following sections provide a model for the dependence structure of ${\bf H}_{r}(t_{k})$, i.e., a parametrization of $\mbf{C}(\bm{\theta}_{\rm lon},\bm{\theta}_{\rm lat})$ in \eqref{eq:tar}. Specifying a valid model for the entire spherical domain that is able to capture global dependence structures is a non-trivial task. However, the following steps rely on the Gaussianity of ${\bf H}_{r}(t_k)$, and hence require to specify only the covariance structure.

%%%%%%%%%%%%%%%%%%%%%%%%%%%%%%%%%%%%%%%%%%%%%%%%%%%%%%%%%%%%%%%%%%%%%%%%
\subsection{Step 2: Longitudinal Structure}\label{second_step}

Here, we focus on $\bm{\theta}_{\rm lon}$, i.e., we provide a model for the dependence structure at different longitudes but at the same latitude. Since the points are equally spaced and on a circle, the implied covariance matrix is circulant under a stationarity assumption \citep{davis1979circulant}, and is more naturally expressed in the spectral domain. The wind behavior on a latitudinal band, however, is not longitudinally stationary. Recently, an evolutionary spectrum approach that allows for changing behavior across large-scale geographical descriptors, was successfully implemented for global annual temperature and wind speed ensembles \citep{castruccio2017evolutionary,jeong2017reducing,cas17b}. Here, we use a similar approach and we model ${H}_{r}(L_{m},\ell_{n},t_{k})$ in the spectral domain via a generalized Fourier transform across longitude. Indeed, if we define $\iota=\sqrt{-1}$ to be the imaginary unit, $c=0, \ldots, N-1$ the wavenumber, then the process can be spectrally represented as
\be
{H}_{r}(L_{m},\ell_{n},t_{k})=\sum_{c=0}^{N-1}f_{L_{m},\ell_{n}}(c)\exp({\iota \ell_{n}c})\widetilde{H}_{r}(c,L_{m},t_{k}), \label{eq:gft}
\ee
with $f_{L_{m},\ell_{n}}(c)$ being a spectrum evolving across longitude, and $\widetilde{H}_{r}(c,L_{m},t_{k})$ the spectral process. 

To better account for the statistical behavior of wind speed, we implement a spatially varying model in which ocean, land, and high mountains above 1,000~m (consistently with \citealp{jeong2017reducing}) are treated as covariates. Therefore $f_{L_{m},\ell_{n}}(c)$ depends on $\ell_{n}$ being in a land, ocean, and high mountain domain, with the following expression:
\be\label{eq:spectra}
f_{L_{m},\ell_{n}}(c)=
\left\{ \begin{array}{ll}
f_{L_{m},\ell_{n}}^{1}(c) & \mbox{if $(L_{m},\ell_{n}) \in {\rm high\ mountain}$}, \\
f_{L_{m},\ell_{n}}^{2}(c)b_{\rm land}(L_{m},\ell_{n};g'_{L_{m}},r'_{L_{m}}) & \mbox{if $(L_{m},\ell_{n}) \in {\rm land}$}, \\
f_{L_{m},\ell_{n}}^{3}(c)\{1-b_{\rm land}(L_{m},\ell_{n};g'_{L_{m}},r'_{L_{m}})\} & \mbox{if  $(L_{m},\ell_{n}) \in {\rm ocean}$},
\end{array} \right.
\ee
where $b_{\rm land}(L_{m},\ell_{n};g'_{L_{m}},r'_{L_{m}})=\sum_{n'=1}^{N}\tilde{I}_{\rm land}(L_{m},\ell_{n};g'_{L_{m}})w(L_{m},\ell_{n}-\ell_{n'};r'_{L_{m}})$ is a smooth function (taper) that allows a transition between the land and the ocean domain. Each of the three components of the spectrum in \eqref{eq:spectra} is parametrized by \citep{castruccio2013global,poppick2014using}:
$|f_{L_{m},\ell_{n}}^{j}(c)|^{2}=\psi_{L_{m},\ell_{n}}^j \{(\alpha^{j}_{L_{m},\ell_{n}})^{2}+4\sin^{2}(c\pi/N) \}^{-\nu_{L_{m},\ell_{n}}^j-1/2}$, for $j=1, 2, 3$, where $(\psi_{L_{m},\ell_{n}}^{j},\alpha_{L_{m},\ell_{n}}^{j},\nu_{L_{m},\ell_{n}}^{j})$ are interpreted as the variance, inverse range, and smoothness parameters, respectively, similarly as for the Mat\'{e}rn spectrum. The parameters are modeled so that their logarithm changes continuously and linearly depends on the altitude, i.e., $\psi_{L_{m},\ell_{n}}^{j}=\beta_{L_{m}}^{j,\psi}\exp[ \tan^{-1}\{A_{L_{m},\ell_{n}} \gamma^{\psi}_{L_{m}}\}]$, $j=1,2$ and $\psi_{L_{m},\ell_{n}}^{3}=\beta_{L_{m}}^{3,\psi}$, where $\beta_{L_{m}}^{j,\psi}>0$, $\gamma^{\psi}_{L_{m}}\in\mathbb{R}$, and $A_{L_{m},\ell_{n}}$ is the altitude at location $(L_{m},\ell_{n})$. Similar notation holds for $\alpha_{L_{m},\ell_{n}}^{j}$ and $\nu_{L_{m},\ell_{n}}^{j}$. Hence, the longitudinal parameters are ${\bm \theta}_{\rm lon}=(\beta_{L_{m}}^{j,\psi},\gamma^{\psi}_{L_{m}},\beta_{L_{m}}^{j,\alpha},\gamma^{\alpha}_{L_{m}},\beta_{L_{m}}^{j,\nu},\gamma^{\nu}_{L_{m}},g'_{L_{m}},r'_{L_{m}})^{\top}$, $j=1,2,3$ and $m=1,\ldots,M$. Since the parameter values for each latitudinal band are independent from other bands, model inference across latitude can be performed independently with distributed computing.

%%%%%%%%%%%%%%%%%%%%%%%%%%%%%%%%%%%%%%%%%%%%%%%%%%%%%%%%%%%%%%%%%%%%%%%%
\subsection{Step 3: Latitudinal Structure}\label{third_step}

We now provide a model for latitudinal dependence, and since the model in \eqref{eq:gft} is independent and identically distributed across $r$ and $t_{k}$, we take out these two indices for simplicity. While \cite{castruccio2013global} and later works have proposed an autoregressive model for $\widetilde{{H}}(c,L_m)$ across $m$ (but independent across $c$), we consider a more general Vector AutoRegressive model of order 1 (VAR(1)) so that $\widetilde{H}(c,L_{m})$ is allowed to also depend on neighboring wavenumbers. We define $\widetilde{\bold{H}}_{L_{m}}=\{ \widetilde{{H}}(1,L_m),\dots,\widetilde{{H}}(N,L_m) \}^{\top}$ and the latitudinal dependence by $\widetilde{\bold{H}}_{L_{m}}=\bm{\varphi}_{L_{m}}\widetilde{\bold{H}}_{L_{m-1}}+{\bold{e}}_{L_{m}}$, 
where ${\bold{e}}_{L_{m}}\stackrel{\rm iid}{\sim}\mathcal{N}(\bold{0},{\bold{\Sigma}_{L_{m}}})$, and $\bm{\varphi}_{L_{m}}$ is a matrix of size $N \times N$ with coefficients of the autoregressive structure across latitude; $\bold{\Sigma}_{L_{m}}$ encodes the dependence for the innovation. To balance flexibility with computational feasibility, we seek a sufficiently sparse but articulated structure for $\bm{\varphi}_{L_{m}}$. We propose a banded diagonally dominant matrix parametrized by $a_{L_{m}}, b_{L_{m}} \in (-1,1)$ for all $m$ values (the explicit expression is available in the supplementary material), $\bold{\Sigma}_{L_{m}}=\text{diag}\{1-\varphi_{L_{m}}(c)^2\}$ and $\varphi_{L_{m}}(c)=\zeta_{L_{m}}\{1+4\sin^{2}(c\pi/N) \}^{-\eta_{L_{m}}}$, where $\zeta_{L_{m}}\in [0,1]$ and $\eta_{L_{m}}>0$ for all $m$. Hence, the latitudinal parameters are ${\bm \theta}_{\rm lat}=(a_{L_m},b_{L_m},\zeta_{L_m},\eta_{L_m})^{\top}$, $m=1,\dots, M$.

\begin{figure}[htb]\centering
\includegraphics[height=2.3in]{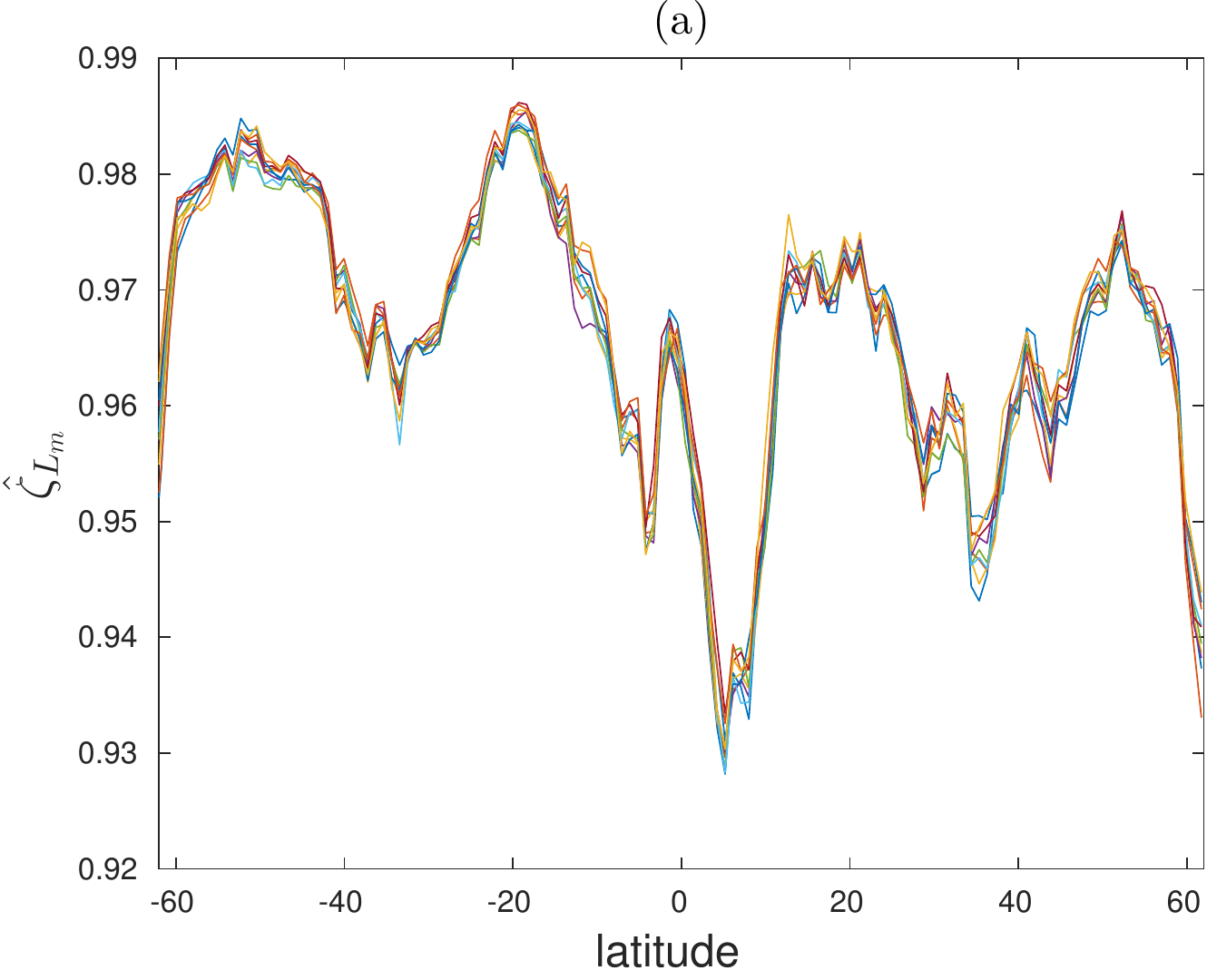}\hspace{0.2cm}
\includegraphics[height=2.3in]{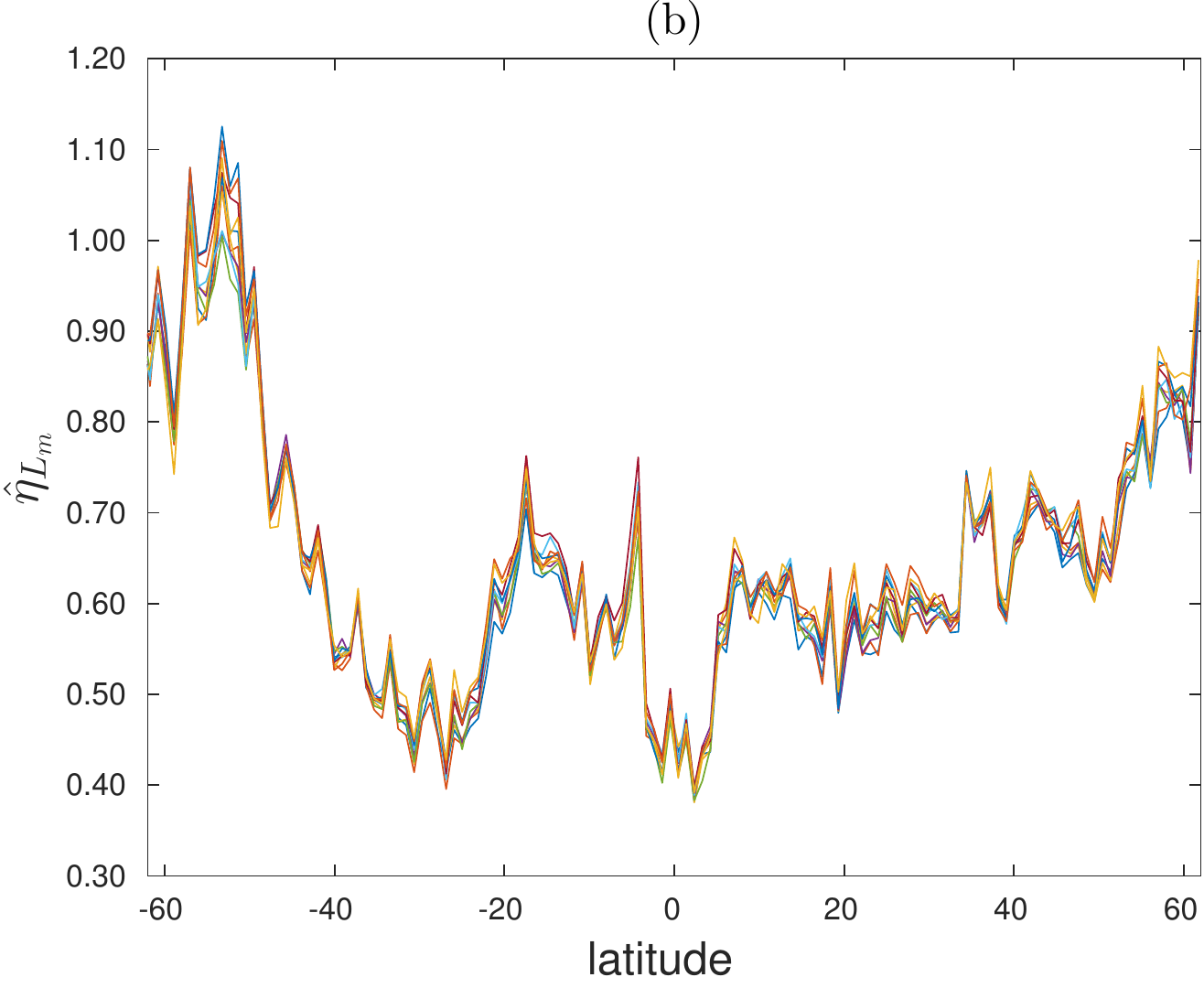}\vspace{0.2cm}
\includegraphics[height=2.3in]{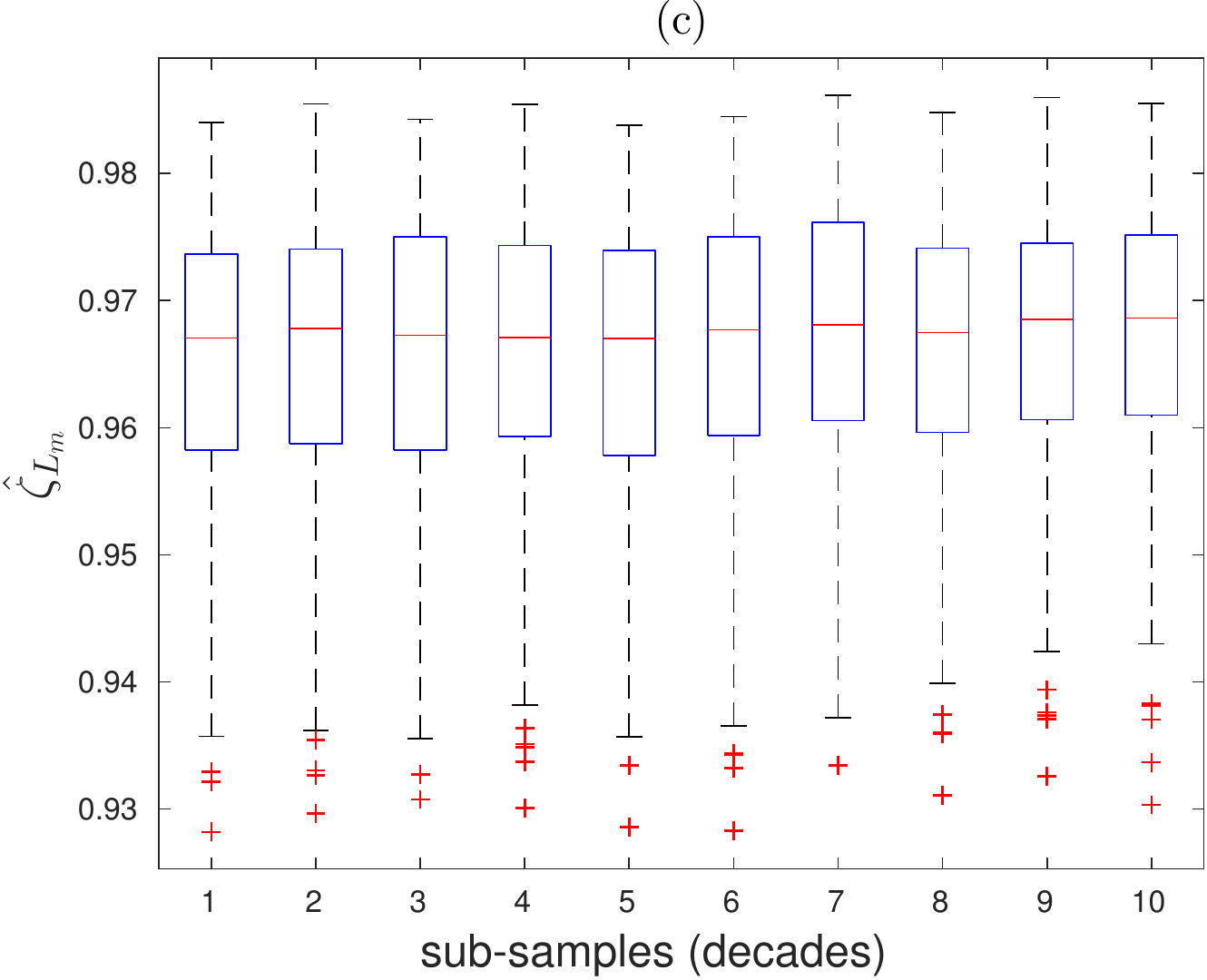}\hspace{0.2cm}
\includegraphics[height=2.3in]{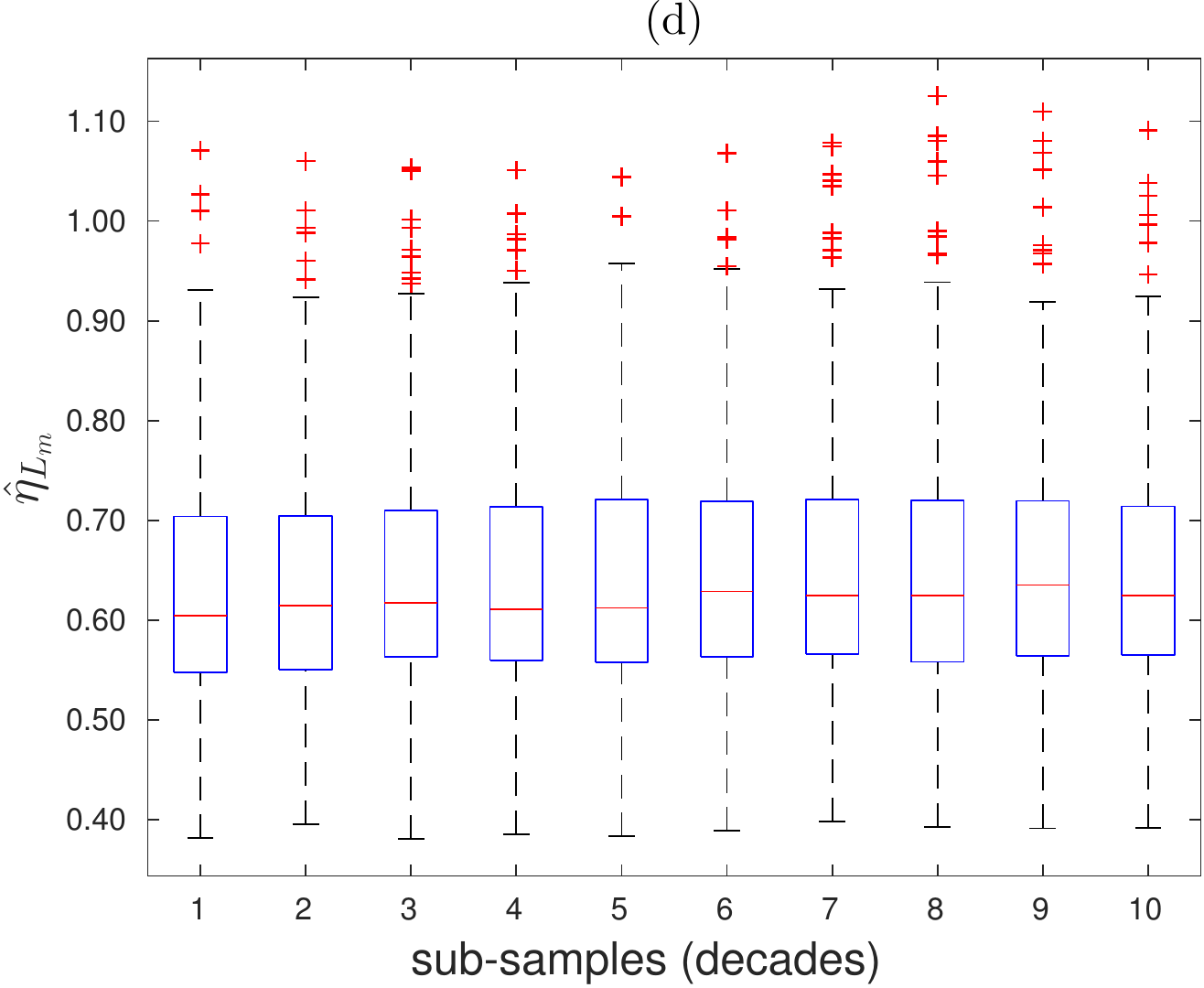}
\caption{Parameter estimates of (a) $\zeta_{L_{m}}$ and (b) $\eta_{L_{m}}$ for 10 sequential sub-samples represented by 10 different colors. Boxplots of estimates of (c) $\zeta_{L_{m}}$ and (d) $\eta_{L_{m}}$ across latitudes at each sub-sample. }\label{fig:subT_alt3_var6N_1}
\end{figure}

We consider ten sequential sub-samples of 95 years (10 years except for the last partition) to reduce the computation burden. The parameter estimates of $\zeta_{L_m}$ and $\eta_{L_m}$ from 10 sub-samples are similar, as shown in Figure~\ref{fig:subT_alt3_var6N_1} (other estimates of longitudinal dependence parameters show similar patterns). Since there is no evidence of a change in latitudinal dependence over time, we consider the average of parameter estimates. Such value is used for combining multiple latitudinal bands and generating surrogates in Section \ref{sec:simul}. $\hat{a}_{L_m}$ and $\hat{b}_{L_m}$ are also shown in Figure~S7.

%%%%%%%%%%%%%%%%%%%%%%%%%%%%%%%%%%%%%%%%%%%%%%%%%%%%%%%%%%%%%%%%%%%%%%%%
\section{Model Comparison}\label{sec:comparison}
%%%%%%%%%%%%%%%%%%%%%%%%%%%%%%%%%%%%%%%%%%%%%%%%%%%%%%%%%%%%%%%%%%%%%%%%

To validate our proposed model based on the Tukey $g$-and-$h$ autoregressive (TGH-AR) process, we compare it with both a Gaussian autoregressive (G-AR) process, and two models with special cases of spatial dependence structure from steps 2 and 3 detailed in Sections~\ref{second_step} and \ref{third_step}. 

\subsection{Comparison with a Gaussian temporal autoregressive process}

\begin{figure}[htb]\centering
\includegraphics[height=3in,width=5.5in]{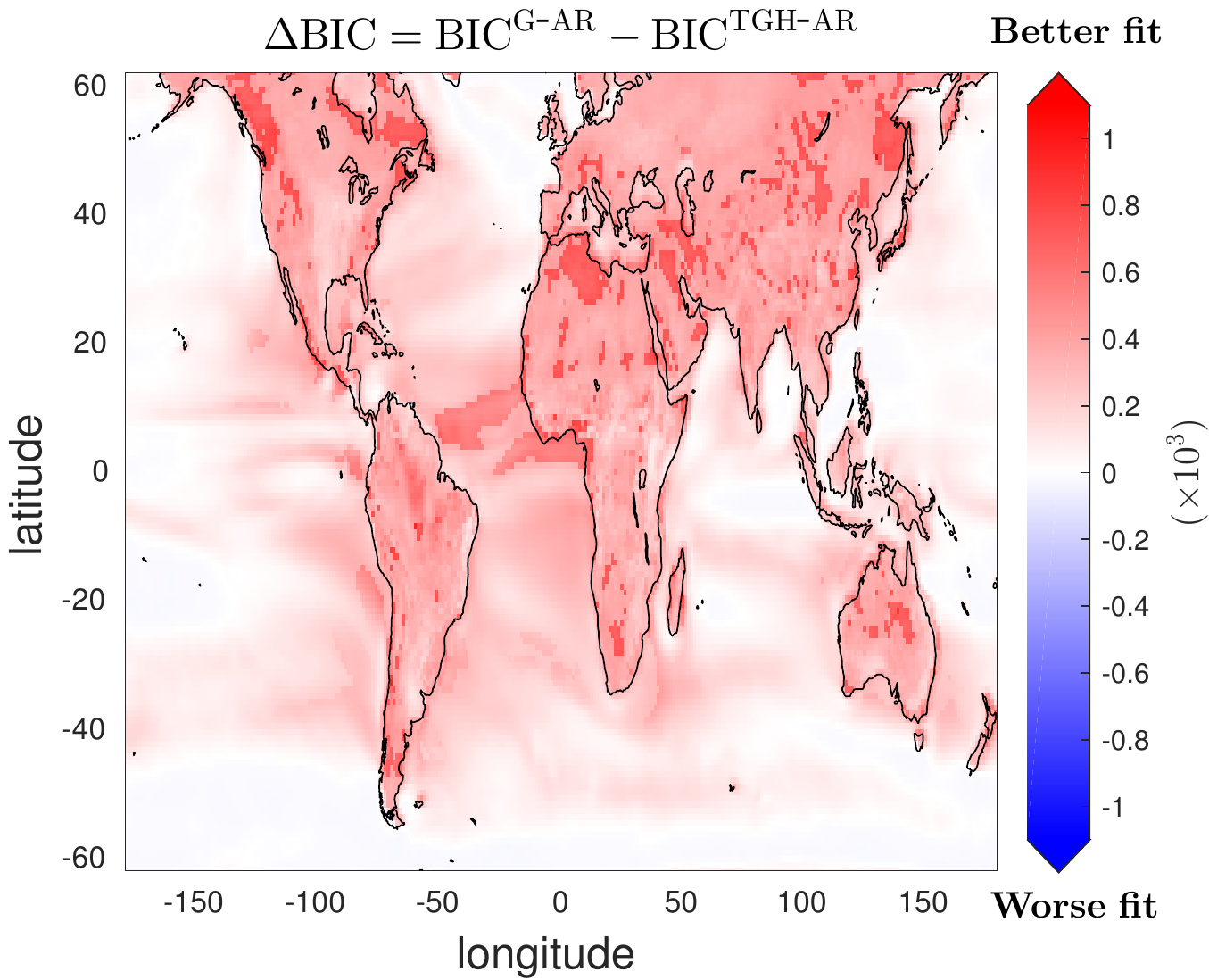}
\caption{Map of differences in BIC between TGH-AR and G-AR from one ensemble member. }\label{fig:diff_AR}
\end{figure}

In our first comparison, we notice that the G-AR process can be obtained from  \eqref{eq_modelT} by assuming ${\bm \xi}=\mbf{0}$, ${\bm \omega}=\mbf{1}$, ${\bf g}=\mbf{0}$, and ${\bf h}=\mbf{0}$; therefore a formal model selection can be performed. Figure~\ref{fig:diff_AR} represents the Bayesian Information Criterion (BIC)  between the two models at each site from one ensemble member. Positive and negative values indicate better and worse model fit of TGH-AR compared to G-AR, respectively. TGH-AR outperforms G-AR in more than $85\%$ of spatial locations, with a considerable improvement in the BIC score (the map scale is in the order of $10^3$). The fit for land sites is overall considerably better for TGH-AR, with peaks in the North Africa area near Tunisia, and in and around Saudi Arabia, in the region of study in Section \ref{sec:simul}. The tropical Atlantic also shows large gains.

\subsection{Comparison with sub-models of global dependence}

The TGH-AR model is also compared with one of no altitude dependence, i.e., where 
\[
\psi_{L_m,\ell_n}^1=\psi_{L_m,\ell_n}^2, \alpha_{L_m,\ell_n}^1=\alpha_{L_m,\ell_n}^2, \nu_{L_m,\ell_n}^1=\nu_{L_m,\ell_n}^2 \implies f_{L_m,\ell_n}^1(c)=f_{L_m,\ell_n}^2(c) 
\]
for all $m, n, c$ in \eqref{eq:spectra}. The model still assumes an evolutionary spectrum with changing behavior across land/ocean \citep{castruccio2017evolutionary}, and is denoted by LAO. We further compare TGH-AR with a model having an autoregressive dependence across latitude, i.e., a model in which $a_{L_m}=b_{L_m}=0$ in the parametrization of $\bm{\varphi}_{L_{m}}$ in Section \ref{third_step}, which we denote as ARL. 

Since both LAO and ARL are special cases of the TGH-AR, a formal comparison of their model selection metrics can be performed (see Table~\ref{tab:metric1}). There is evidence of a considerable improvement from LAO to ARL, hence the need to incorporate the altitude while modeling the covariance structure. The additional,  smaller (although non-negligible, as the BIC improvement is approximately $10^5$) improvement from ARL to TGH-AR underscores the necessity of a flexible model that is able to account for dependence across both wavenumbers and latitude. 

\begin{table}[htb]\centering
\caption{\label{tab:metric1} Comparison of the number of parameters (excluding the temporal component), the normalized restricted log-likelihood, and BIC for three different models: LAO, ARL, and TGH-AR. The general guidelines for $\Delta$loglik$/\{NMK(R-1)\}$ are that values above 0.1 are considered to be large and those above 0.01 are modest but still sizable \citep{castruccio2013global}.}\vspace{0.3cm}
\begin{tabular}{|c||c|c|c|c|}
\hline
Model  &  LAO &  ARL & TGH-AR\\
\hline
\# of parameters & $1338$ & $2142$ & $2408$\\  
\hline
$\Delta$loglik$/\{NMK(R-1)\}$ & $0$ & $0.0440$ & ${\bf 0.0443}$ \\
\hline
BIC ($\times 10^8)$ &  $-5.8963$ & $-6.0511$ & ${\bf -6.0521}$\\
\hline
\end{tabular}
\end{table}

All three models can also be compared via local contrasts, as the residuals in \eqref{eq:innovation} are approximately Gaussian. We focus on the contrast variances to assess the goodness of fit of the model in terms of its ability to reproduce the local dependence \citep{jun2008nonstationary}:
\be\label{delta_con}
\begin{split}
\Delta_{ew;m,n}&=&\frac{1}{KR}\sum_{k=1}^{K}\sum_{r=1}^{R}\{{H}_{r}(L_{m},\ell_{n},t_{k})-{H}_{r}(L_m,\ell_{n-1},t_{k})\}^{2},\\
\Delta_{ns;m,n}&=&\frac{1}{KR}\sum_{k=1}^{K}\sum_{r=1}^{R}\{{H}_{r}(L_{m},\ell_{n},t_{k})-{H}_{r}(L_{m-1},\ell_{n},t_{k})\}^{2},
\end{split}
\ee
where $\Delta_{ew;m,n}$ and $\Delta_{ns;m,n}$ denote the east-west and north-south contrast variances, respectively. 

\begin{table}[b]\centering
\caption{\label{tab:diff}$25$th, $50$th and $75$th percentiles of two difference metrics over ocean, land, and high mountain near the Indian ocean.}\vspace{0.3cm}
\begin{tabular}{|c||c||c|c|c|}
\hline
Metric  & Region & $25$th & $50$th & $75$th \\
\hline
\multirow{3}{*}{$[ \{ \Delta_{ew;m,n}-\hat{\Delta}_{ew;m,n}^{\rm ARL} \}^{2} - \{ \Delta_{ew;m,n}-\hat{\Delta}_{ew;m,n}^{\rm TGH-AR} \}^{2} ]\times 10^4$ }  & ocean & $0$ & $0$ & $0$ \\
& land & $-14$ & $0$ & $16$ \\
& mountain & $-8$ & $5$ & $22$ \\
\hline
\multirow{3}{*}{$[ \{ \Delta_{ns;m,n}-\hat{\Delta}_{ ns;m,n}^{\rm ARL} \}^{2} - \{ \Delta_{ ns;m,n}-\hat{\Delta}_{ ns;m,n}^{\rm TGH-AR} \}^{2} ]\times 10^4$} & ocean & $-1$ & $1$ & $2$ \\
& land & $-2$ & $2$ & $11$ \\
& mountain & $-2$ & $1$ & $7$\\
\hline
\end{tabular}
\end{table}

We compare ARL with TGH-AR, and compute the squared distances between the empirical and fitted variances. We find that the TGH-AR shows a better model fit in the case of the north-south contrast variance but that there is no noticeable difference between the two models in the case of the east-west variances. A representation of these differences for the small region of interest near South Africa ($13.75^{\circ}{\rm E}\sim 48.75^{\circ}{\rm E}$ and $30^{\circ}{\rm S}\sim 4^{\circ}{\rm N}$) is given in Figure~S8. Positive values are obtained when TGH-AR is a better model fit than the ARL; negative values are obtained when ARL is the better model fit. Figure~S8(a) and (b) show that dark red colors are more widely spread over mountains, and that no clear difference is shown over the ocean. Results presented in Table~\ref{tab:diff} are consistent with the visual inspection and the two metrics, in particular, over mountain areas, show larger values than those obtained for the ocean areas. In a global mean or median of the metrics, there is no significant difference between the two models.

%%%%%%%%%%%%%%%%%%%%%%%%%%%%%%%%%%%%%%%%%%%%%%%%%%%%%%%%%%%%%%%%%%%%%%%%
\section{Generation of Stochastic Surrogates}\label{sec:simul}
%%%%%%%%%%%%%%%%%%%%%%%%%%%%%%%%%%%%%%%%%%%%%%%%%%%%%%%%%%%%%%%%%%%%%%%%

Once the model is properly defined and validated, we apply it to produce surrogate runs and train the SG with $R=5$ climate runs. A comprehensive sensitivity analysis on the number of elements in the training set can be found in \cite{jeong2017reducing}. We use the SG to obtain an assessment of the uncertainty of monthly wind power density, and compare it with the results of the full extent of the LENS runs. 

The mean structure of the model is obtained by smoothing the ensemble mean $\overline{\bf W}$, but such estimate is highly variable. For each latitude and longitude (i.e., each $n$ and $m$),  we fit a spline $\widetilde{W}(L_{m},\ell_{n},t_{k})$ which minimizes the following function \citep{castruccio2017evolutionary,jeong2017reducing}: $\lambda \sum_{k=1}^{K}\Big\{\overline{W}(L_{m},\ell_{n},t_{k})-{\widetilde{W}}(L_{m},\ell_{n},t_{k})\Big\}^{2}+(1-\lambda)\sum_{k=1}^{K} \Big\{\nabla_2 \widetilde{W}(L_{m},\ell_{n},t_{k}) \Big\}^{2}$, $\nabla_2$ being the discrete Laplacian. We impose $\lambda=0.99$ to give significant weight to the spline interpolant in order to reflect the varying patterns of monthly wind fields over the next century. For each spatial location, harmonic regression of a time series may also be used to estimate the mean structure, but for the sake of simplicity, we opt for a non-parametric description.
Once $\bm{\theta}=(\bm{\theta}_{\rm Tukey}^{\top},\bm{\theta}_{\rm space-time}^{\top})^{\top}$ is estimated from the training set, surrogate runs can be easily generated by the Algorithm~\ref{alg1}. \vspace{.3cm}

\begin{spacing}{1.5}
\begin{algorithm}[t!]
\caption{Generate surrogates}\label{alg1}
\begin{algorithmic}[1] 
\Procedure{Generate surrogates}{}  \\
Generate ${\bold{e}}_{L_{m}}\stackrel{\rm iid}{\sim}\mathcal{N}(\bold{0},{\bold{\Sigma}}_{L_{m}})$ as in Section \ref{third_step}.\\
Compute the VAR(1) process $\widetilde{\bold{H}}_{L_{m}}$ as in Section \ref{third_step}.\\
Compute ${H}_{r}(L_{m},\ell_{n},t_{k})$ from \eqref{eq:gft}\\
Compute ${\bm \epsilon}_{r}$ with equation \eqref{eq:tar}, and obtain $\widetilde{\bf D}_{r}$ from the Tukey $g$-and-$h$ transformation \eqref{eq_modelT}\\
Obtain the SG run as $\widetilde{\bold W}+\widetilde{\bf D}_{r}$, where\newline
$
\widetilde{\bf W}=\{\widetilde{W}(L_{1},\ell_{1},t_{1}),\dots,\widetilde{W}(L_{M},\ell_{1},t_{1}),\widetilde{W}(L_{1},\ell_{2},t_{1}),\dots,\widetilde{W}(L_{M},\ell_{N},t_{K})  \}^{\top}.
$
\EndProcedure
\end{algorithmic}
\end{algorithm}
\end{spacing}

%\begin{itemize}
%\itemsep0em
%\item[] Step 1. Generate ${\bold{e}}_{L_{m}}\stackrel{\rm iid}{\sim}\mathcal{N}(\bold{0},{\bold{\Sigma}}_{L_{m}})$ as in \eqref{eq:var1};
%\item[] Step 2. Compute $\widetilde{\bold{H}}_{L_{m}}$ with expressions \eqref{eq:var1};
%\item[] Step 3. Compute ${H}_{r}(L_{m},\ell_{n},t_{k})$ with expression \eqref{eq:gft};
%\item[] Step 4. Compute ${\bm \epsilon}_{r}$ with equation \eqref{eq:tar}, and obtain $\widetilde{\bf D}_{r}$ from the Tukey $g$-and-$h$ transformation \eqref{eq_modelT};
%\item[] Step 5. Obtain the SG run as $\widetilde{\bold W}+\widetilde{\bf D}_{r}$, where 
%\[
%\widetilde{\bf W}=\{\widetilde{W}(L_{1},\ell_{1},t_{1}),\dots,\widetilde{W}(L_{M},\ell_{1},t_{1}),\widetilde{W}(L_{1},\ell_{2},t_{1}),\dots,\widetilde{W}(L_{M},\ell_{N},t_{K})  \}^{\top}.
%\]
%\end{itemize}

We generate forty SG runs with the model presented in this work and compared them with the original forty LENS runs. As clearly shown in Figures~\ref{fig:statistics}(a) and S9(a), the ensemble means from the training set and the SG runs are visually indistinguishable.

We also evaluate both models in terms of structural similarity index; to that end, we compare local patterns of pixel intensities that have been standardized for luminance and contrast (Figure~S10) \citep{wang2004image,cas17}. We observe that the SG runs from the Tukey $g$-and-$h$ case produce maps that are visually more similar to the original LENS runs than those in the Gaussian case (see also Figure~S12 for the measures of skewness and kurtosis).

\begin{figure}[htb]\centering
\includegraphics[height=2in,width=3.2in]{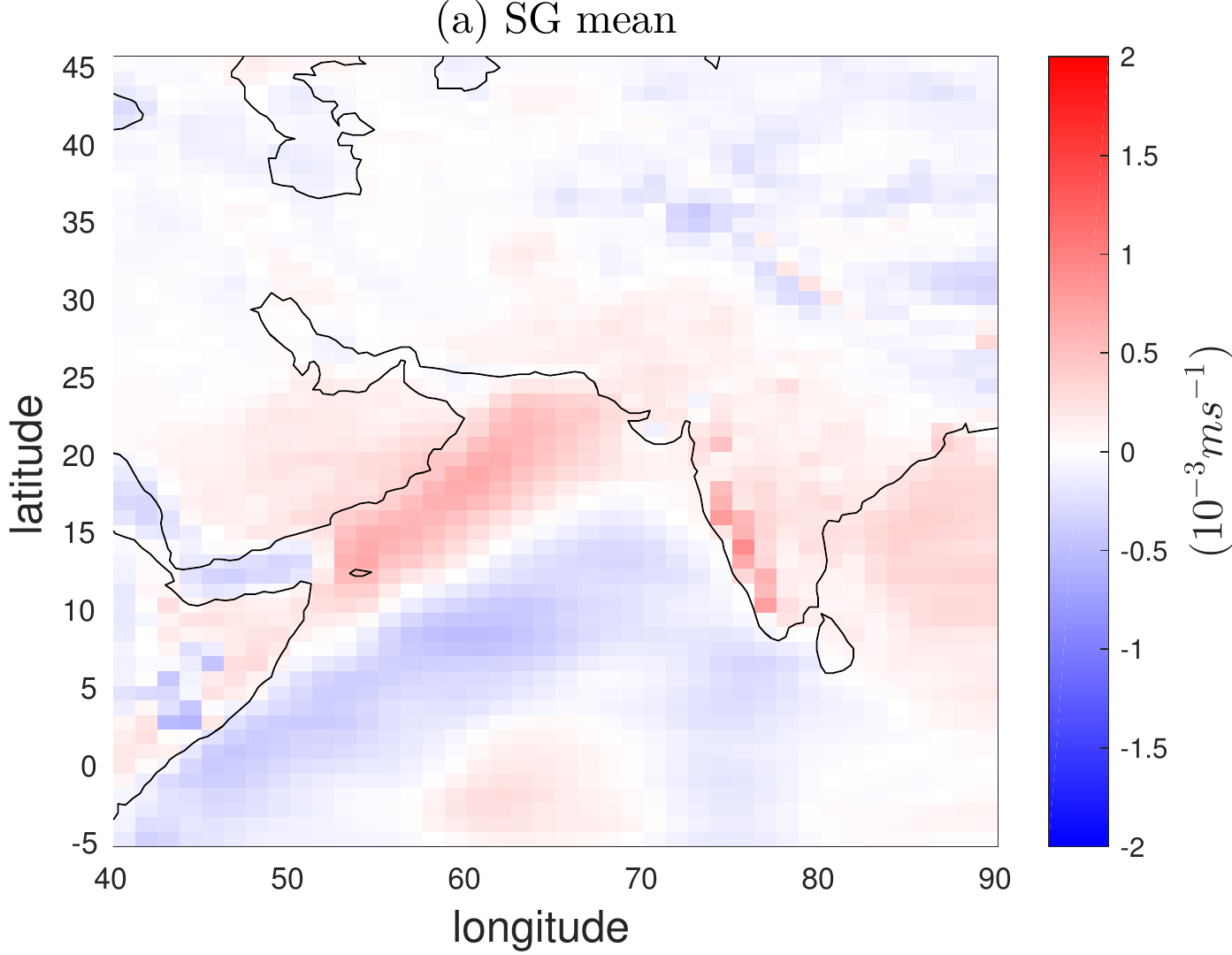}\hspace{0.2cm}
\includegraphics[height=2in,width=3.2in]{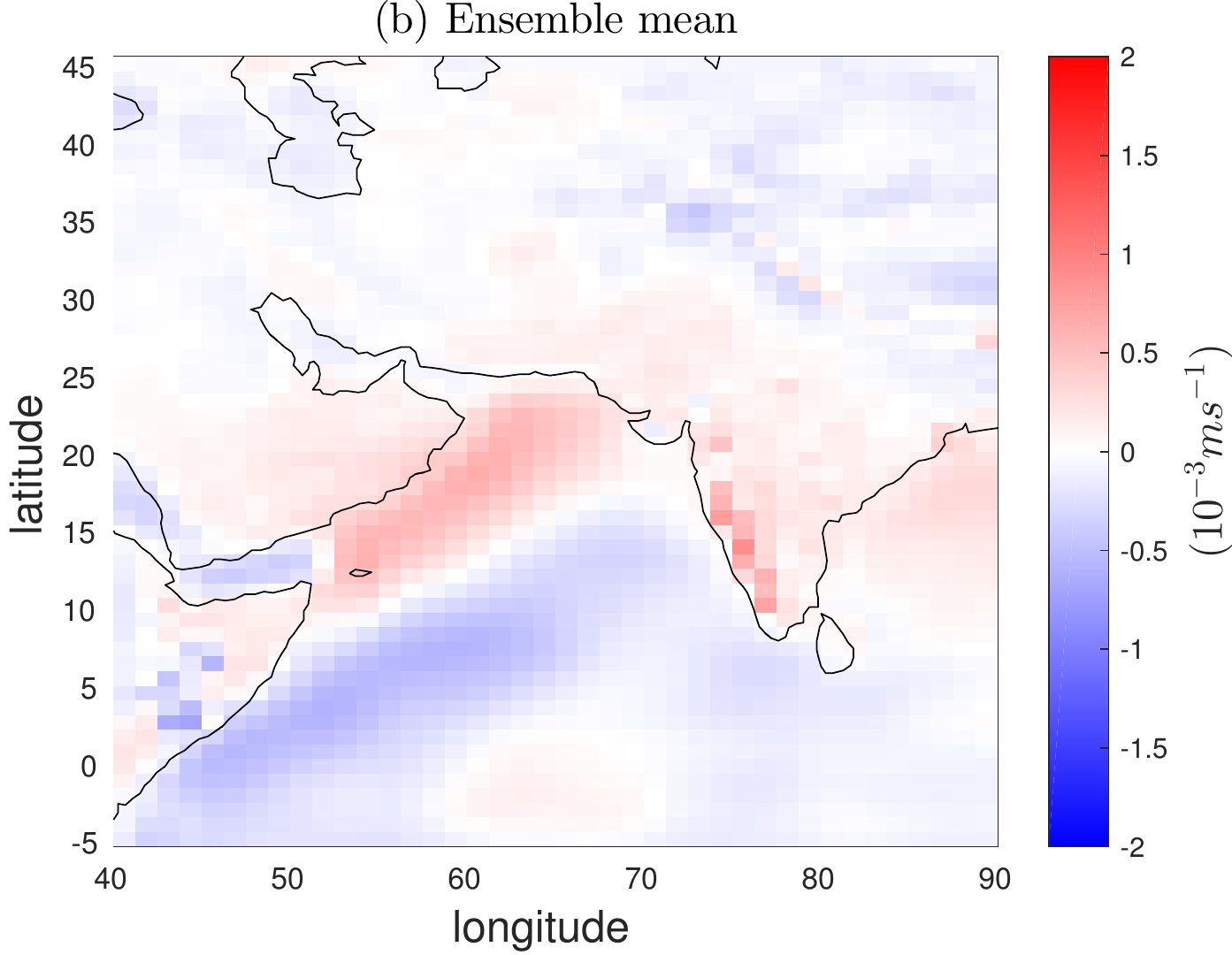}\vspace{0.2cm}
\includegraphics[height=2in,width=3.2in]{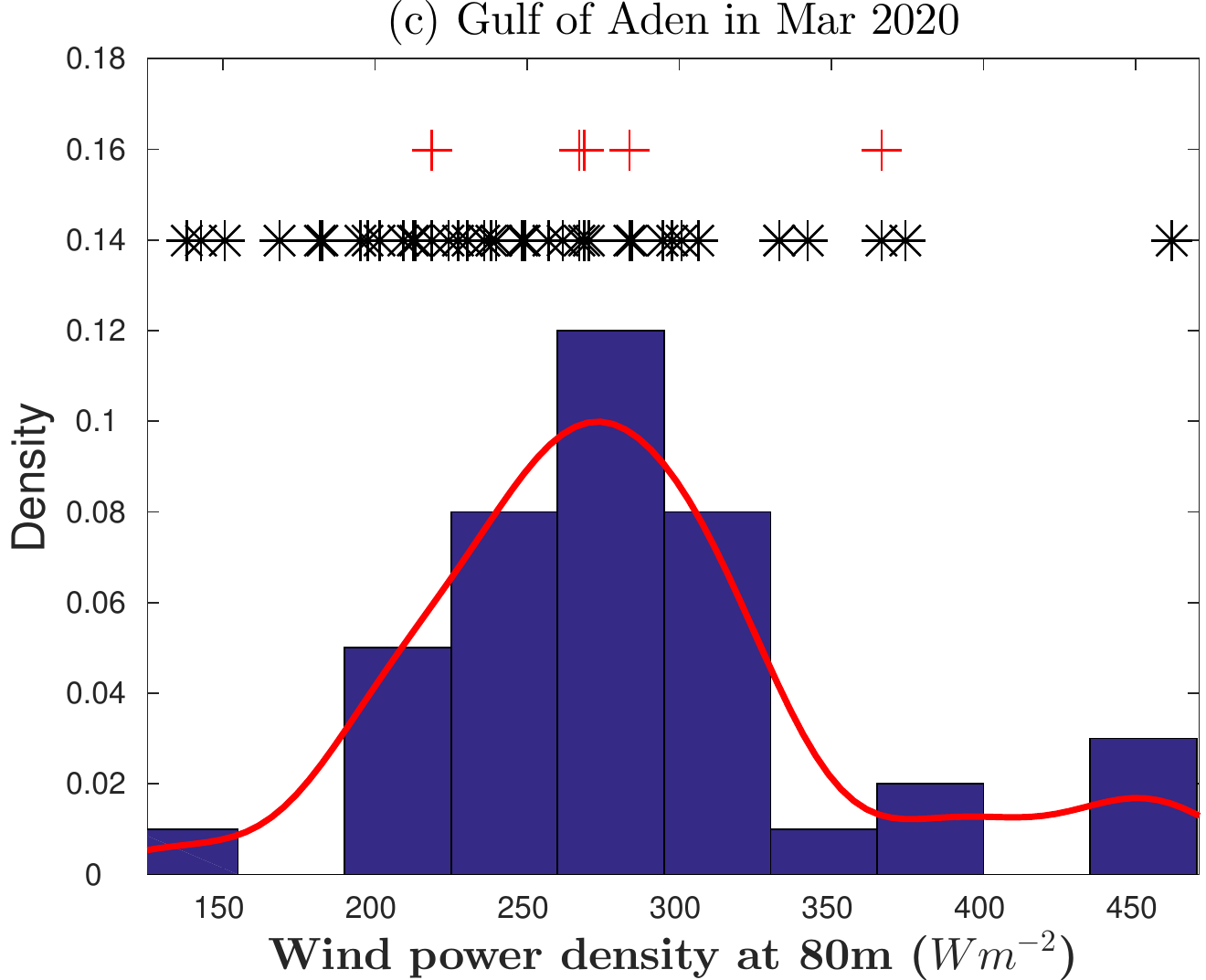}\hspace{0.2cm}
\includegraphics[height=2in,width=3.2in]{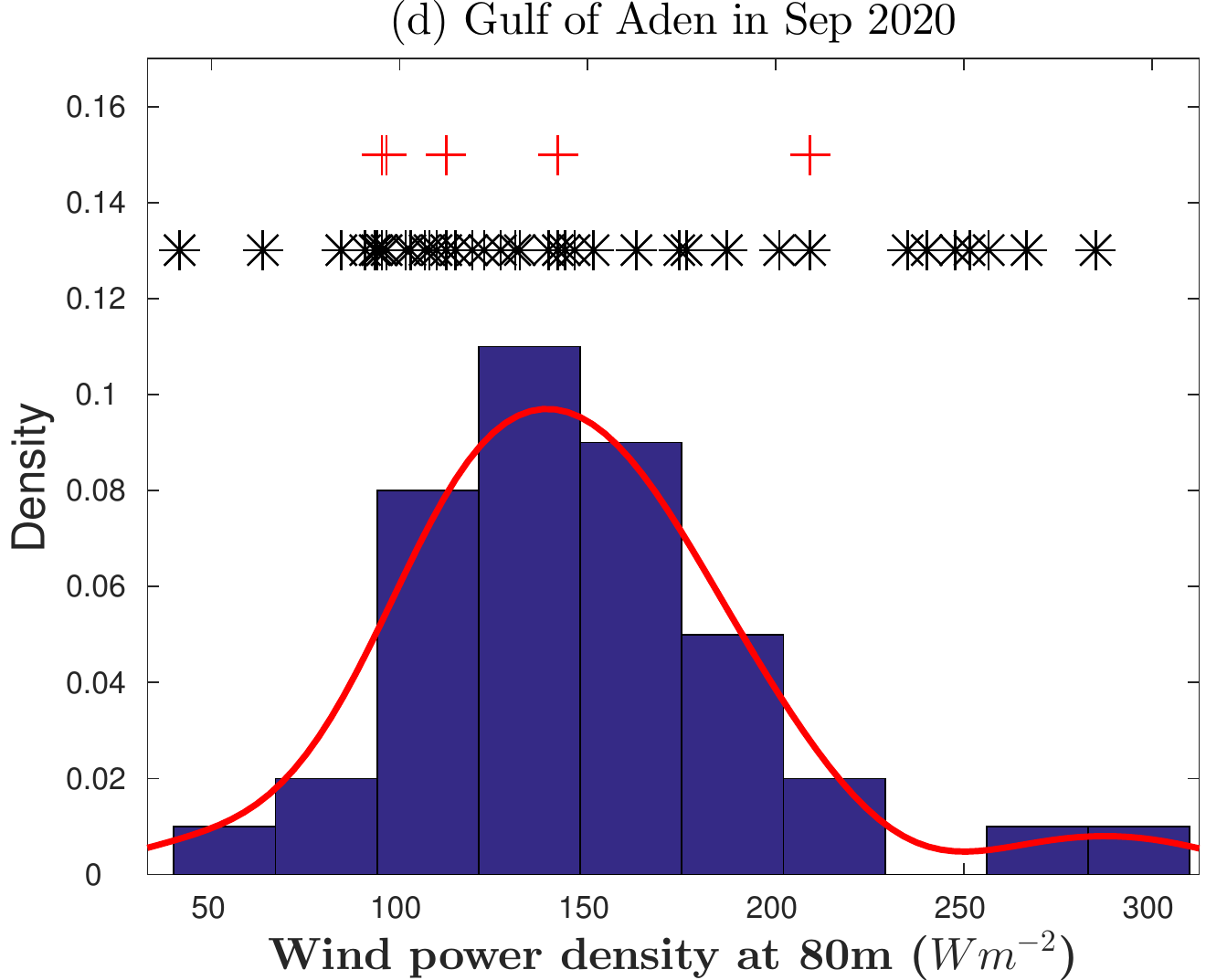}
\caption{Maps of (a) the mean from SG runs and (b) the ensemble mean from the near-future ($2013\textendash 2046$) near-surface wind speed trends near the Indian ocean. Histogram of the distribution of the wind power density at $80$~m with nonparametric density in red for the forty SG runs near the  Gulf of Aden (c) in March 2020, and (d) in September 2020 ($\ast$ represents the LENS runs, + represents the five LENS runs in the training set of the SG). }\label{fig:near_future_slope}
\end{figure}

We further compare LENS and SG in terms of near-future trend (2013$\textendash$2046), a reference metric for the LENS \citep{kay2015community} that was used to illustrate the influence of the internal variability on global warming trends. We compute near-future wind speed trends near the Indian ocean for each of the SG and LENS runs. Results are shown in Figure~\ref{fig:near_future_slope}(a) and (b). One can clearly see that the mean near-future wind trends by the SG runs are very similar to those from the training set of LENS runs. 

We subsequently provide an assessment of the wind energy potential. The wind power density (WPD) (in $W m^{-2}$) evaluates the wind energy resource available at the site for conversion by a wind turbine. WPD can be calculated as $\text{WPD}=0.5\rho u^{3}, u=u_{r}(z/z_{r})^{\alpha}$, where $\rho$ is the air density ($\rho=1.225$ kgm$^{-3}$ in this study), $u$ is the wind speed at a certain height $z$, $u_{r}$ is the known wind speed at a reference height $z_{r}$, and $\alpha=1/7$ \citep{peterson1978use,newman2013extrapolation}. We focus our analysis on the Gulf of Aden ($46.25^{\circ}$E and $12.72^{\circ}$N), a narrow channel connecting the Red Sea to the Indian Ocean with high wind regimes \citep{yip2017high}, and we choose to work on WDP at 80 meters, a standard height for wind turbines \citep{holt2012trends,yip2017high}, in the year 2020. 

Results for March and September are represented in Figure~\ref{fig:near_future_slope}(c) and (d), with the histogram representing the SG runs, a superimposed estimated nonparametric density in red, and the LENS runs on top with an asterisk marker. Both histograms have rightly skewed distributional shapes, as also reflected by the distribution of the entire LENS. It is clear that the distribution resulting from the SG runs is more informative than the five LENS runs in the training set (see red cross markers on top), and matches the uncertainty generated by the forty LENS runs. Figure~S10 also reports the time series of five LENS runs against five SG runs. 

\begin{figure}[htb]\centering
\includegraphics[height=2.25in,width=4.5in]{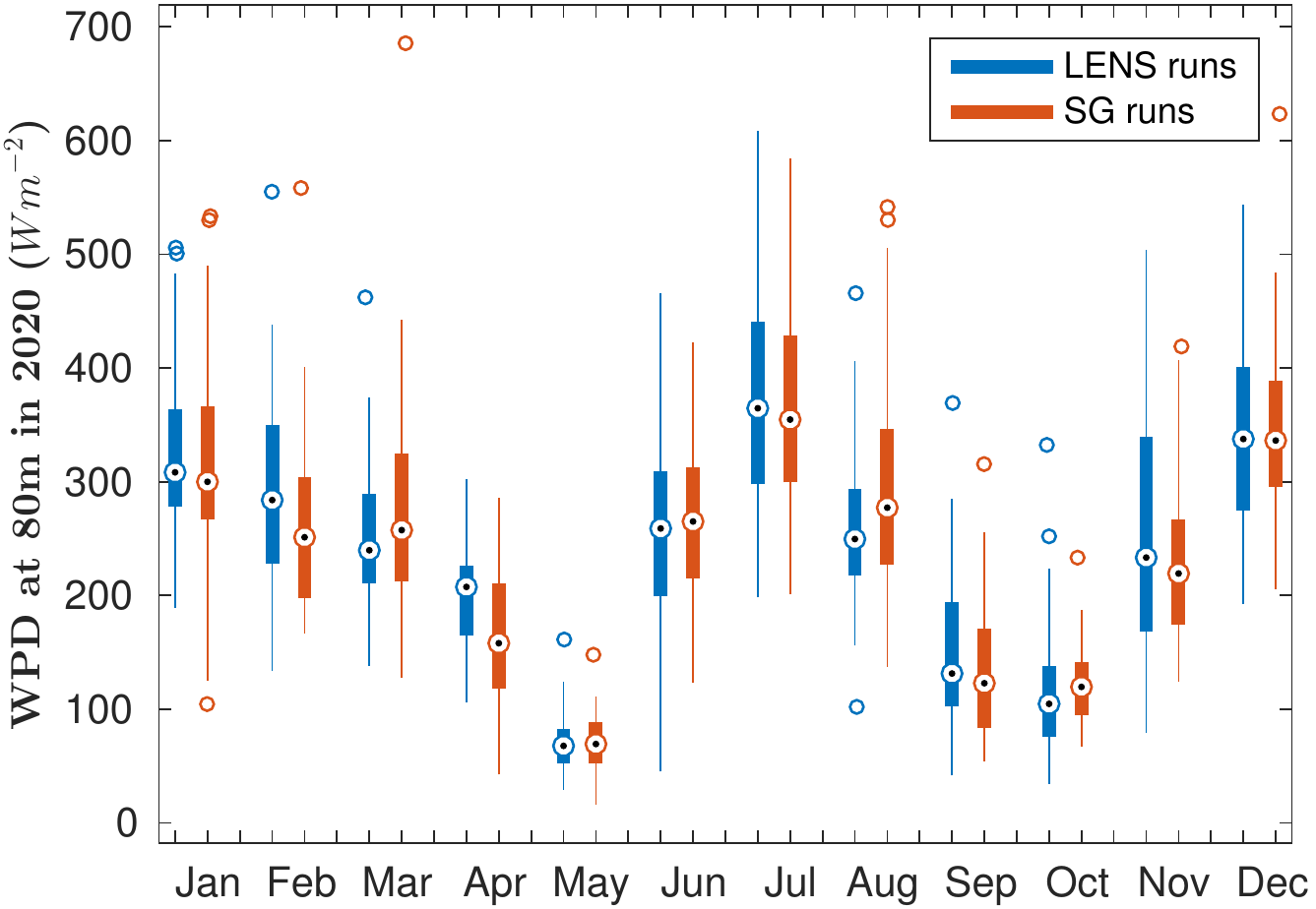}
\caption{Boxplots of the distribution of the wind power density at $80$~m, in $2020$, for 40 LENS runs and the 40 SG runs based on the Tukey $g$-and-$h$ case near the Gulf of Aden. }\label{fig:WPD_2020_boxplots2}
\end{figure}

In Figure~\ref{fig:WPD_2020_boxplots2}, we report the boxplots of the distribution of WPD in 2020 for the LENS against the SG runs across all months. The point estimates and ranges of the WPD values from the LENS runs are well-matched by those from the SG, with a slight misfit in April and November. The importance of such results cannot be understated: the SG is able to capture the interannual WPD patterns, as well as its internal variability in a region of critical importance for wind farming. The internal variability in the months of high wind activity such as July is such that the WPD can be classified from fair to very high according to standard wind energy categories \citep{archer2003spatial}, and the SG can reproduce the same range with as little as five runs in the training set.

%%%%%%%%%%%%%%%%%%%%%%%%%%%%%%%%%%%%%%%%%%%%%%%%%%%%%%%%%%%%%%%%%%%%%%%%
\section{Discussion and Conclusion}\label{sec:concl}
%%%%%%%%%%%%%%%%%%%%%%%%%%%%%%%%%%%%%%%%%%%%%%%%%%%%%%%%%%%%%%%%%%%%%%%%
 
In this work, we propose a non-Gaussian, multi-step spectral model for a global space-time data set of more than 220 million points. Motivated by the need for approximating a computer output with a faster surrogate, we provide a fast, parallelizable and scalable methodology to perform inference on a massive data set and to assess the uncertainty of global monthly wind energy. 

Our proposed model relies on a trans-Gaussian process, the Tukey $g$-and-$h$, which allows controlling skewness and tail behavior with two distinct parameters. This class of models is embedded in a multi-step approach to allow inference for a nonstationary global model while also capturing site-specific temporal dependence, and it clearly outperforms currently available Gaussian models. 

Our model has been applied as a Stochastic Generator (SG), a new class of stochastic approximations that assesses more efficiently the internal variability for wind energy resources in developing countries with poor observational data coverage, using global models. Our results suggest that the uncertainty produced by the SG with a training set of five runs is very similar to that from the forty LENS runs in regions of critical interest for wind farming. Therefore, our model can be used as an efficient surrogate to assess the variability of wind energy at the monthly level, a clear improvement from the annual results presented by \cite{jeong2017reducing}, and an important step forward towards the use of SGs at policy-relevant time scales.

Given the ubiquity of big data in complex structures evolving in space and time, it is natural to incorporate  scalable algorithms from other disciplines into our modeling framework. To handle extremely large covariance matrices, we can consider the integration of high-performance computing in exact likelihood inference and predictions \citep{abdulah2017exageostat} and approximations through parallel Cholesky decompositions of $\mathcal{H}$-matrices \citep{hackbusch1999sparse,hackbusch2015hierarchical} on different architectures \citep{litvinenko2017likelihood}.

%One significant benefit of the proposed model in generating runs is that the proposed model can produce smaller and bigger extreme values and describe skewed distribution in time compared to the Gaussian models. Another benefit is that SG can provide a simple, but efficient tool for producing many surrogate runs and is useful to assess the uncertainty from internal variability without relying on expensive computing power and time.  
 
% {\color{red} i) More bigger data with parallel computing and advanced library of matrix calculation such as ExaGeoStat. ii) More flexible model for multivariate wind components such as $U$ and $V$ components (eastward and northward wind components), and even $\omega$ components (vertical components). iii) Three dimensional covariance functions on spheres for wind components at different pressures or altitudes?}

\begin{spacing}{1.3}
\bibliography{papers1}
\bibliographystyle{Chicago}
\end{spacing}

\end{document}